\documentclass[lettersize,journal]{IEEEtran} 
\usepackage[mathscr]{euscript}
\usepackage{color}
\usepackage{optidef}
\usepackage[algo2e,ruled,vlined]{algorithm2e}
\usepackage{float}
\usepackage{graphicx}
\usepackage{tikz,pgfplots}
\usepackage{upgreek}
\usepackage{multirow}
\usepackage{yfonts}
\usepackage{mathtools}
\usepackage[normalem]{ulem}
\usepackage{latexsym}
\usepackage{url}
\usepackage{amsmath,amssymb,amsbsy,amsfonts}
\usepackage{mathrsfs}
\usepackage{bbm}
\usepackage[export]{adjustbox}

\newcommand\restr[2]{{
  \left.\kern-\nulldelimiterspace 
  {#1}\vphantom{\big|} \right|_{#2}}}

\DeclareMathOperator*{\argmin}{argmin} 

\newcommand{\R}{\mathbb{R}}

\newcommand{\U}{\mathbb{U}}

\newcommand{\D}{\mathcal{D}}

\newcommand{\J}{\mathcal{J}}

\newcommand{\Reg}{\mathcal{R}}

\newcommand{\ipar}{f}

\newcommand{\noise}{\bs n}
\newcommand{\data}{ \bs d }

\newcommand{\bs}[1]{\ensuremath{\boldsymbol{#1}}}

\usepackage{xspace}

\newcommand{\LI}{\mathcal{L}}
\renewcommand{\H}{\mathcal{H}}

\newcommand{\rr}{\ensuremath{\boldsymbol r}}

\usepackage{cite}
\usepackage{amsmath,amssymb,amsfonts,amsthm}
\usepackage{graphicx}
\usepackage{textcomp}
\def\BibTeX{{\rm B\kern-.05em{\sc i\kern-.025em b}\kern-.08em
    T\kern-.1667em\lower.7ex\hbox{E}\kern-.125emX}}
\usepackage[math]{cellspace}
\usepackage{graphicx}
\usepackage{algorithm}
\usepackage{algpseudocode}

\usepackage{float}
\usepackage{enumerate}
\usepackage{url}
\usepackage{multirow}

\newtheorem{rem}{Remark}
\newtheorem{prop}{Proposition}

\usepackage{wrapfig,lipsum,booktabs}

\usepackage[switch,columnwise]{lineno}

\pgfplotsset{compat = 1.17}
\begin{document} 
\title{ProxNF: Neural Field Proximal Training for High-Resolution 4D Dynamic Image Reconstruction}

\author{Luke Lozenski, \IEEEmembership{Graduate Student Member, IEEE},  Refik Mert Cam, \IEEEmembership{Graduate Student Member, IEEE},\\
Mark D. Pagel,
Mark A. Anastasio, \IEEEmembership{Fellow, IEEE}, and Umberto Villa \IEEEmembership{Member, IEEE}
\thanks{This work was supported in part by the NIH under grants R44 OD023029, R01 EB031585, \& R01 EB034261, and in part by the Joint Center for Computational Oncology of the University of Texas System, Oncological Data and Computational Sciences program. LL also acknowledges support from the Imaging Science Pathway fellowship funded by the NIH under grant T32 EB014855.}
\thanks{Luke Lozenski is with the Department of Electrical and Systems Engineering,
 Washington University in St$.\,$Louis, St$.\,$Louis, MO 63130, USA.}
 \thanks{Refik Mert Cam is with the Department of Electrical and Computer Engineering, University of Illinois Urbana-Champaign, Urbana, IL 61801, USA}
\thanks{Mark D Pagel is with the Department of Medical Physics at the University of Wisconsin–Madison, Madison, WI 53705, USA.}
\thanks{Mark Anastasio is with the Department of Bioengineering, University of Illinois Urbana-Champaign, Urbana, IL 61801, USA.}
\thanks{Umberto Villa is with the Oden Institute for Computational Engineering and Sciences, University of Texas at Austin, Austin, TX 78712.}
\thanks{Further author information: (Send correspondence to Umberto Villa.)\\ E-mail: uvilla@oden.utexas.edu, Telephone: +1 512 232-3453}}

\markboth{Journal of \LaTeX\ Class Files,~Vol.~14, No.~8, August~2021}%
{Shell \MakeLowercase{\textit{et al.}}: A Sample Article Using IEEEtran.cls for IEEE Journals}


\maketitle
\vspace{-.11cm}
\begin{abstract}

Accurate spatiotemporal image reconstruction methods are needed for a wide range of biomedical research areas but face challenges due to data incompleteness and computational burden. Data incompleteness arises from the undersampling often required to increase frame rates, while computational burden emerges due to the memory footprint of high-resolution images with three spatial dimensions and extended time horizons. Neural fields (NFs), an emerging class of neural networks that act as continuous representations of spatiotemporal objects, have previously been introduced to solve these dynamic imaging problems by reframing image reconstruction as a problem of estimating network parameters. Neural fields can address the twin challenges of data incompleteness and computational burden by exploiting underlying redundancies in these spatiotemporal objects. This work proposes ProxNF, a novel neural field training approach for spatiotemporal image reconstruction leveraging proximal splitting methods to separate computations involving the imaging operator from updates of the network parameters. 
Specifically, ProxNF evaluates the (subsampled) gradient of the data-fidelity term in the image domain and uses a fully supervised learning approach to update the neural field parameters. 
This method is demonstrated in two numerical phantom studies 
and an in-vivo application to tumor perfusion imaging in small animal models using dynamic contrast-enhanced photoacoustic computed tomography (DCE PACT).

\end{abstract}



\section{Introduction}

Dynamic image reconstruction with high spatial and temporal resolution is important for understanding the evolution of many biological systems \cite{taruttis2012fast, stern1993dynamic,gamper2008compressed}. In many dynamic image reconstruction problems, measurements are undersampled and frame-by-frame reconstruction methods yield unsatisfactory accuracy. In such cases, higher accuracy can be achieved by utilizing a spatiotemporal reconstruction method in which the image is treated as a single spatiotemporal object \cite{Liang2007}. However, large-scale spatiotemporal image reconstruction can quickly become computationally prohibitive and exceed memory requirements in three spatial dimensions. Therefore, high-resolution dynamic image reconstruction needs to exploit underlying redundancies in the image to reduce the number of  parameters necessary to represent the image \cite{kortman1967redundancy, liang1994efficient}.  

On the one hand, optimization-based image reconstruction that employs semiseparable or low rank approximations has often been advocated to address these challenges. Previous work has demonstrated the use of such approaches to enable several dynamic imaging applications, including X-ray tomography  \cite{cai2014cine}, positron electron tomography \cite{wernick1999fast}, magnetic resonance imaging\cite{haldar2010spatiotemporal,liang1994efficient}, and more recently also photoacoustic computed tomography \cite{cam2023low, cam2024spatiotemporal,cam2023dynamic}. On the other hand, the use of machine learning, and neural fields in particular, has been actively investigated and shows great promise for the solution of this class of image reconstruction problems. Neural fields (NFs) are a particular class of neural networks that provide a continuous representation of an object in space and time. Using NFs reduces a dynamic image reconstruction problem into estimating network weights by solving a nonconvex optimization problem \cite{LozenskiNeuralFields, ReedKimAnirudhetal2021, choromanska2015loss}. Unlike other deep learning methods, NFs are self-supervised. The NF model is trained only using the measurement data and does not require ground truth images, or labels\cite{jaiswal2020survey}. Additionally, NFs allow for memory-efficient representation of the object \cite{dupont2021coin, MartelLindellLinetal2021}  and provide implicit regularization based on the neural architecture\cite{SitzmannMartelBergmanetal2020}. 

Previous works with NFs utilized first-order (accelerated) gradient-based methods \cite{MartelLindellLinetal2021, SitzmannMartelBergmanetal2020, XieTakikawaSaitoEtAl21, LiuSunetal2021, TancikPratuletal20, dupont2021coin, shen2022nerp}, such as Adam \cite{KingmaBa2014}, to train the network parameters. However, training an NF for image reconstruction in this fashion requires the imaging operator and its adjoint to be computed at each gradient update, greatly slowing down training \cite{LozenskiNeuralFields, SunLiuXieEtAl2021, ReedKimAnirudhetal2021}. Furthermore, while computationally efficient implementations of discrete-to-discrete imaging operators are sometimes available, the implementation of a continuous-to-discrete imaging operator acting on an NF representation incurs severe computational and memory burden due to the need for repeated evaluation of the NF \cite{LozenskiNeuralFields}.

This work proposes ProxNF, a proximal splitting method for self-supervised training of an NF  to enable memory-efficient, high-resolution spatiotemporal image reconstruction.  The key contribution of this splitting method is a separation of concerns between the updates of  NF parameters and calls to the imaging operator. In this approach, an  update direction in the image domain is computed using the gradient of the data-fidelity term, which requires an evaluation of the discrete-to-discrete imaging operator and its adjoint. Once this update direction is formed, no further computations involving the imaging operator are needed. The NF parameters are then updated independently from the imaging operator by solving a supervised subproblem (the proximal operator). In essence, this method reframes the original self-supervised problem as a sequence of supervised subproblems while reducing the number of calls to the imaging operator and thus the computational burden. Furthermore, the update direction can be computed in a memory-efficient manner via randomized evaluation of the data-fidelity term by stochastically subsampling the imaging frames at each iteration. ProxNF allows for reconstruction of large, high-resolution spatiotemporal images that were previously intractable.  This method is demonstrated by use of numerical phantom studies  in two and three spatial dimensions and an in-vivo  imaging experiment of tumor perfusion using dynamic contrast-enhanced photoacoustic tomography (DCE PACT).

Furthermore, the general idea behind the proposed method, i.e. splitting computation involving the imaging operator from the training of the neural network representing the object, is readily extendable to other classes of neural networks, such as deep image priors (DIP, \cite{ulyanov2018deep}). In the field of medical imaging, in particular, DIPs have been applied to x-ray computed tomography (CT) \cite{baguer2020computed}, positron electron tomography (PET) \cite{gong2018pet}, and dynamic magnetic resonance imaging \cite{yoo2021time}. The DIP approach has several similarities to NFs. These include 1) the use of a neural network to represent the sought-after object, 2) the implicit regularization provided by such representation, and 3) the self-supervised approach to train the network parameters for a specific input (measurement data) by the solution of an inverse problem. The main difference between NFs and DIPs lies in the choice of architecture. While NFs use a (often dense) neural network architecture mapping spatiotemporal coordinates to image intensities at a specific location, DIPs use a deep convolutional neural network to map a random but fixed code vector to the sought-after image/object estimate.  Notably, an expectation maximization algorithm attuned for PET imaging was proposed in \cite{gong2018pet,li2022neural,li2023pet} that leverages the alternating direction method of multiplier (ADMM, \cite{boyd2011distributed}) to split computations involving the imaging operator to the training of the neural representation of the sought-after object. 
Algorithmically, the proposed method differs from previous methods because ProxNF uses a forward-backward splitting approach while  \cite{gong2018pet,li2022neural,li2023pet} employ backward-backward splitting (ADMM).
Advantages of the proposed ProxNF are: 1) The use of an NF representation reduces the memory footprint for large-scale high-dimensional images; 2) The introduction of a stochastic approximation of the data fidelity term increases computational efficiency and further reduce memory requirements.

The remainder of this paper is structured as follows.  In Section \ref{sec:background}, spatiotemporal image reconstruction is formulated as an optimization problem and  reconstruction methods utilizing NFs for representing the dynamic object are introduced. In Section \ref{sec:method}, the proposed ProxNF method for spatiotemporal image reconstruction is derived and analyzed. This section also discusses a specific NF architecture, the partition of unity network, and introduces an adaptation to this architecture for representing high-resolution, high-dimensional dynamic images. In Section \ref{sec:numerical_studies}, the design of two numerical studies is presented along with the computational framework for virtual DCE PACT imaging of a numerical tumor perfusion phantom in two and three spatial dimensions.  
Section \ref{sec:results} presents the results of these numerical studies. In Section \ref{sec:in-vivo}, an application to in-vivo DCE PACT imaging of tumor perfusion in small animal models is presented.
Section \ref{sec:conclusion} presents the conclusions drawn from these results and future extensions.

\section{Background}
\label{sec:background}
\subsection{Spatiotemporal image reconstruction}
\subsubsection{Continuous-to-discrete formulation}
Image reconstruction aims to estimate an object $\ipar$ based on a set of indirect measurements $\data$. The object $\ipar$ belongs to an infinite-dimensional Hilbert space $\U$. The measurements $\data$ belong to a finite-dimensional space $\D$ and are associated to $\ipar$ via a possibly nonlinear imaging operator $\H:\U \rightarrow \D$, also known as a forward operator, via \begin{equation}\label{eqn:imaging_eq}
    \data = \H(\ipar) + \noise,
\end{equation}
where $\noise$ represents additive noise. 

 This work is concerned with the specific case of dynamic image reconstruction. That is, $\ipar$ is a spatiotemporal function belonging to $\U = \mathbb{L}^2(\Omega_T)$ with $\Omega_T = \Omega \times  [0,T]$, where $\Omega\subset \R^d$ ($d = 2$ or 3) denotes the spatial domain and $T$ is the time horizon. 
 
 In dynamic image reconstruction, the imaging operator $\H$ is the concatenation of imaging operators $\{\H_k\}_{k=1}^K$, one for each imaging frame $k=1, \ldots K$, where $K$ denotes the number of frames. Under the quasi-static approximation that the object $f$ does not change during the acquisition of the $k$-th imaging frame, the measurements $\data_k \in \mathcal{D}_k$ collected at the $k$-th frame are described by 

\begin{equation}
\label{eqn:imaging_eq_frame}
    \data_k = \H_k(\ipar(\cdot, t_k)) + \noise_k, \quad k=1, \ldots, K,
\end{equation} 
where $t_k = (k-1)\Delta T\in[0,T]$ is the sampling time of frame $k$, $\Delta T$ is the sampling interval, $\ipar(\cdot,t_k)$ is the object at frame $k$, and $\noise_k$ represents measurement noise. For several applications of practical relevance, the measurement aperture translates or rotates between imaging frames \cite{Liang2007, sidky2006accurate, velikina2015reconstruction, haldar2011low, cam2023dynamic, cam2024spatiotemporal}, yielding imaging operators $\{\H_k\}_{k=1}^K$ that are different for each frame and measurements that are severely undersampled. In what follows, the notation
\begin{equation}
\begin{array}{l}
\displaystyle \data  \coloneqq \sum_{k=1}^K \data_k\otimes \mathbf{e}_k, \quad \noise   \coloneqq \sum_{k=1}^K \noise_k\otimes \mathbf{e}_k, \\
\displaystyle \H(\ipar)   \coloneqq  \sum_{k=1}^K \H_k(\ipar(\cdot,t_k)) \otimes \mathbf{e}_k 
\end{array}
\label{eq:cc_spatiotemporal}
\end{equation}
is introduced to express the dynamic imaging model in Eqn. \eqref{eqn:imaging_eq_frame} in the general form of Eqn. \eqref{eqn:imaging_eq}. Above, $\{\mathbf{e}_k\}_{k=1}^K$ indicates the canonical basis of $\mathbb{R}^K$ and the symbol $\otimes$ represents the outer product of two vectors yielding a rank-1 matrix.

Using the above notation, spatiotemporal image reconstruction can be formulated as an optimization problem

 \begin{equation}\label{eqn:Objective}
    \min_{\ipar} \J(\ipar) := \LI(\ipar)  + \Reg(\ipar).
 \end{equation}
 
 Above, the cost functional $\J:\U \rightarrow \R$ is the sum of the data-fidelity functional $\LI:\U\rightarrow \R$, which is minimized when $\H(\ipar) = \data$, and the regularization functional $\Reg:\U \rightarrow \R$, which encodes prior information on the sought-after image. In this work, the data-fidelity functional $\LI(\ipar) =\sum_{k=1}^K \LI_k(\ipar(\cdot, t_k))$ is chosen as 
\begin{equation}
\begin{split}   \LI(\ipar) &\coloneqq \frac{1}{2\sigma^2} \| \H(\ipar) - \data\|^2   \\
 \displaystyle     &=  \sum_{k=1}^K \frac{1}{2\sigma^2}\| \H_k(\ipar(\cdot, t_k)) - \data_k\|^2,
\end{split}
\label{eq:continuous_data_fidelity}
\end{equation}
which corresponds to the negative log-likelihood of observing the data $\data$ given the object estimate $\ipar$ under the assumption of additive Gaussian measurement noise $\boldsymbol{n} \sim \mathcal{N}(\boldsymbol{0}, \sigma^2 \boldsymbol{I})$.

\subsubsection{Discrete-to-discrete formulation}
To derive the discrete-to-discrete formulation, the spatial domain $\Omega$ is discretized using a Cartesian grid with $M$ non-overlapping voxels. The coordinates of the voxel centers are denoted by the vectors $\{\rr_m\}_{m=1}^M$. The spatial and temporal expansion functions $\beta_m(\rr)$ and $\gamma_k(r)$ are chosen as the indicator functions of the $m$-th spatial voxel ($m=1,\ldots,M$) and the $\left(t_k - \frac{1}{2}\Delta T, t_k + \frac{1}{2}\Delta T\right)$ time interval ($k=1,\ldots,K$), respectively.
The finite-dimensional subspace $\mathbb{U}_{M,K} \subset \mathbb{U}$ consists of all functions $f_{M,K}$ of the form
\begin{equation}\label{eqn:disc_decomp}
    f_{M,K}(\rr,t) = \sum_{k=1}^K \sum_{m=1}^M {\rm f}_{m,k} \beta_m(\rr) \gamma_k(t),
\end{equation}
where ${\rm f}_{m,k} \in \R$ ($m=1,\ldots,M$, $k=1,\ldots,K$) are the expansion coefficients. Eqn. \eqref{eqn:disc_decomp} defines an isomorphism $\mathcal{I}: \R^{M \times K} \mapsto \mathbb{U}_{M,K}$ mapping the matrix $\mathbf{F} $ with entries $\{{\rm f}_{m,k}\}_{m,k}$ to the function $f_{M,K}(\rr,t)$.

The inner product in $\mathbb{U}_{M,K}$ is defined as
\begin{equation}
\label{eqn:inner-product}
\begin{array}{ll}
\displaystyle (f_{M,K}, g_{M,K})_{\mathbb{U}_{M,K}} & \coloneqq \int_{\Omega_T} f_{M,K}(\rr, t)g_{M,K}(\rr, t) d\rr dt \\[1mm]
\displaystyle & = \sum_{k=1}^K\sum_{m=1}^M {\rm f}_{m,k} \, {\rm g}_{m,k} V_{m,k},
\end{array}
\end{equation}
where ${\rm g}_{m,k} $ are the expasion coefficients of $g_{M,K}$ and $V_{m,k} = \int_{\Omega_T}  \beta^2_m(\rr) \gamma^2_k(t) d\rr\, dt$ denotes the volume of the $(m,k)$-th spatiotemporal voxel.
The orthogonal projection operator $\Pi_{M,K}: \mathbb{U} \mapsto \R^{M,K}$ maps an object function $f \in \U$ to the matrix $\mathbf{F}$ with entries
$$ {\rm f}_{m,k} = \frac{1}{V_{m,k}}\int_{\Omega_T} f(\rr, t) \beta_m(\rr) \gamma_k(t) d\rr\, dt, \quad \begin{array}{l}
m=1,\ldots,M,\\
k=1,\ldots,K.
\end{array}$$
The adjoint operator $\Pi_{M,K}^*: \R^{M,K} \mapsto \mathbb{U}_{M,K} \subset \mathbb{U}$ is defined with respect to the inner product in Eqn. \eqref{eqn:inner-product} as
\begin{equation}
\label{eqn:adjoint_projection}
\Pi_{M,K}^*(\mathbf{F}) \coloneqq \sum_{k=1}^K\sum_{m=1}^M \frac{1}{V_{m,k}}{\rm f}_{m,k} \beta_m(\rr) \gamma_k(t).
\end{equation}

The notations $\mathbf{f}_k$ and $\Pi_{M,K}^k(f)$ are also introduced to denote $k$-th column of $\mathbf{F}$ and $\Pi_{M,K}(f)$, respectively. The adjoint operator $\left(\Pi_{M,K}^k\right)^*: \R^{M} \mapsto \mathbb{U}_{M,K} \subset \mathbb{U}$ is then defined as
\begin{equation}
\label{eqn:adjoint_projection_frame}
\begin{split}
\left(\Pi_{M,K}^k\right)^*(\mathbf{f}_k) &\coloneqq \sum_{m=1}^M \frac{1}{V_{m,k}}{\rm f}_{m,k} \beta_m(\rr) \gamma_k(t) \\
{} &= \Pi_{M,K}^*(\mathbf{f}_k \otimes \mathbf{e}_k).
\end{split}
\end{equation}
 
The discrete-to-discrete imaging operators $\{\mathbf{H}_k:\R^{M} \rightarrow \mathcal{D}_k \}_{k=1}^K$ are constructed by evaluating their continuous-to-discrete counterpart acting on the spatial expansion functions $\{ \beta_m(\rr) \}_m$. Specifically, they are defined as
\begin{equation}
 \mathbf{H}_k \coloneqq \sum_{m=1}^M \mathcal{H}_k(\beta_m(\cdot)) \otimes \mathbf{i}_m,
 \end{equation}
where $\{ \mathbf{i}_m \}_{m=1}^M$ denotes the canonical basis of $\R^M$.
The discrete-to-discrete spatiotemporal imaging operator $\mathbf{H}: \R^{M\times K} \mapsto \D $ is defined in a similar manner to its continuous-to-discrete counterpart, c.f. Eqn. \eqref{eq:cc_spatiotemporal}. Its action on a matrix $\mathbf{F}$ representing the discretized spatiotemporal object function $f$ is expressed as
\begin{equation}
\mathbf{H}(\mathbf{F}) = \sum_{k=1}^K \left(\mathbf{H}_k \mathbf{f}_k\right) \otimes \mathbf{e}_k.
\end{equation}

Using the above notation, the discrete counterpart of the optimization-based formulation of spatiotemporal image reconstruction is given by
\begin{equation}
\min_{\mathbf{F} \in \R^{M\times K}} \mathbf{J}(\mathbf{F})  \coloneqq \mathbf{L}(\mathbf{F})+ \mathbf{R}(\mathbf{F}),
\end{equation}
where
$$ \mathbf{L}(\mathbf{F}) = \sum_{k=1}^K\mathbf{L}_k(\mathbf{f}_k) 
= \frac{1}{2\sigma^2} \sum_{k=1}^K \| \mathbf{H}_k \mathbf{f}_k - \data_k\|^2
$$
and
$$ \mathbf{R}(\mathbf{F}) = \Reg(\mathcal{I}(\mathbf{F})).$$

\subsection{Neural fields}

A neural field (NF) is a neural network that provides a representation of continuous objects, such as images, videos, and audio signals \cite{XieTakikawaSaitoEtAl21}. An NF can be described as  a neural network $\Phi_{\boldsymbol \xi}(\boldsymbol{x}):\Omega_T \rightarrow \R$  that maps every spatiotemporal point $\boldsymbol{x}=(\boldsymbol{r}, t) \in \Omega_T$ to the values of an image at that point. NF architectures represent a finite-dimensional, nonlinear topological subspace of $L^2(\Omega_T)$ parametrized by a set of network parameters $\boldsymbol \xi \in \R^{N_p}$. 
For many classes of images, in particular biomedical images\cite{shen2022nerp, juhl2021implicit,molaei2023implicit}, NFs can provide accurate  representations using a fixed network architecture and requiring a relatively small number of parameters $N_p$ compared to the number of voxels needed to represent the image with a comparable resolution \cite{dupont2021coin}. NFs are then able to significantly reduce the memory required for storing high-resolution, high-dimensional images. 

Given a target function $f \in \mathbb{L}^2(\Omega_T)$, an NF can be trained to represent $f$ by solving the supervised embedding problem

\begin{equation}\label{eqn:embedding}
    \min_{\boldsymbol \xi} \frac{1}{2}\| f- \Phi_{\boldsymbol \xi}\|_2^2,
\end{equation}
where $\| f- \Phi_{\boldsymbol \xi}\|_2$ denotes the $L^2(\Omega_T)$ norm of $f- \Phi_{\boldsymbol \xi}$.
 Once trained, the NF can be evaluated at arbitrary points in $\Omega_T$ and, thus, rendered at any chosen resolution. 

\subsubsection{Spatiotemporal image reconstruction using neural fields}\label{sec:nf_dynamic_imaging}

In the case of spatiotemporal image reconstruction, the object $f$ is not directly accessible.  Previous works have demonstrated that NFs can be directly trained using measurement data alone by replacing the sought-after object $f$ with its NF representation $\Phi_{\boldsymbol \xi}$ into the optimization problem  in Eq. \eqref{eqn:Objective}\cite{LozenskiNeuralFields, ReedKimAnirudhetal2021}. This gives rise to the self-supervised training problem 
\begin{equation}\label{eqn:INR_obj}
    \min_{\boldsymbol \xi}  \J(\Phi_{\boldsymbol \xi}) \coloneqq 
    \frac{1}{2\sigma^2} \| \H(\Phi_{\boldsymbol \xi}) - \data\|^2 + \Reg(\Phi_{\boldsymbol \xi}).
\end{equation}

Advantages of using an NF for spatiotemporal image reconstruction in Eq. \eqref{eqn:INR_obj} include reduced memory requirements for optimization. However, this does come with a few disadvantages. First, optimizing over the space of network parameters  to train the NF is a nonconvex optimization problem, even in the case when the original problem in Eq. \eqref{eqn:Objective} is convex. The nonconvex nature of this problem means there are no guarantees on finding a global minimizer; additionally, non-convex problems are in general more difficult to solve than convex problems. Second, while computationally efficient implementations of several canonical discrete-to-discrete imaging operators are readily available, the evaluation of a continuous-to-discrete imaging operator acting on the NF representation, $\H(\Phi_{\boldsymbol \xi})$ is computationally expensive and further slows down training.

\subsubsection{Partition of unity networks}

One particular NF network architecture is the partition of unity network (POUnet) \cite{LeeTraskPateletal21}, which has previously been utilized for spatiotemporal image reconstruction in two spatial dimensions \cite{LozenskiNeuralFields}. The  POUnet consists of three components: a trainable partition of unity $\boldsymbol \Psi_{\boldsymbol{\eta}}:\Omega_T \rightarrow [0,1]^P$ parameterized by a set of weights $\boldsymbol{\eta} \in \R^W$, a trainable coefficient matrix $\boldsymbol C\in \R^{L\times P}$, and a fixed set of spatiotemporal basis functions $\boldsymbol B:\Omega_T \rightarrow \R^L$.  Specifically, the POUnet NF representation is given by 
\begin{equation}
\label{eq:pounet_def}
\Phi_{(\boldsymbol{\eta},\boldsymbol C)}(\boldsymbol{x}) = \boldsymbol \Psi_{\boldsymbol{\eta}}^T(\boldsymbol{x}) \boldsymbol C^T \boldsymbol B(\boldsymbol{x}),\quad \forall \boldsymbol{x} = (\boldsymbol{r},t) \in \Omega_T,
\end{equation}
where the partition of unity network $\boldsymbol \Psi_{\boldsymbol{\eta}}(\boldsymbol{x}) = (\psi_{\boldsymbol{\eta}}^1(\boldsymbol{x}),\hdots,\psi_{\boldsymbol{\eta}}^P(\boldsymbol{x}))$ is such that
$$ \sum_{p=1}^P \psi^p_{\boldsymbol{\eta}}(\boldsymbol{x}) = 1, \text{ and } \psi^p_{\boldsymbol{\eta}}(\boldsymbol{x}) \geq 0. $$

By letting $\boldsymbol{\xi} = (\boldsymbol{\eta}, \boldsymbol{C})$, an approximate solution to the minimization problem in Eq. \eqref{eqn:INR_obj} can be found using an alternating minimization strategy

\begin{equation}\label{eqn:alt_min}
\begin{array}{c}
     \boldsymbol C^{i+1} = \argmin_{\boldsymbol{C}} \mathcal{J}( \Phi_{(\boldsymbol \eta^i,\boldsymbol C)}) \\ 
     \boldsymbol{\eta}^{i+1} = \argmin_{\boldsymbol{\eta}}\mathcal{J}(\Phi_{(\boldsymbol{\eta}, \boldsymbol C^{i+1})}).
\end{array}   
\end{equation}

 The advantage of using the above alternating optimization strategy is that, due to the linearity of $\Phi_{(\boldsymbol \eta,\boldsymbol C)}$ with respect to $\boldsymbol C$, the first minimization problem admits a unique global minimum when $\mathcal{J}$ is strongly convex. Thanks to this property, the POUnet representation possesses theoretically guaranteed approximation properties based on a priori error estimates for piecewise polynomial interpolation \cite{LeeTraskPateletal21}.

\section{Method}
\label{sec:method}

This section presents the main contribution of this work: the development of a proximal training method for NF-based spatiotemporal image reconstruction, ProxNF. In the proposed method, all computations involving the imaging operator are separated from updates of the NF parameters. Furthermore, ProxNF leverages preexisting discrete implementations of the imaging operator and fast NF embedding methods to scale to high-resolution, high-dimensional imaging problems. This work also develops a modified partition of unity NF architecture that is capable of  representing dynamic images with three spatial dimensions.

\subsection{ProxNF: a proximal training method for neural field dynamic image reconstruction}
Previously, it has been shown that NFs can be utilized for small-scale spatiotemporal image reconstruction problems and trained with gradient-based methods\cite{LozenskiNeuralFields, ReedKimAnirudhetal2021}. However, these gradient-based methods become very slow when scaling to high-resolution, high-dimensional problems. This lack of scalability is due to the  difficulties outlined in Section \ref{sec:nf_dynamic_imaging} of nonconvexity and expensive imaging operator evaluations.

Notably, these disadvantages are greatly reduced in the supervised embedding problem in Eq. \eqref{eqn:embedding} for a few reasons. First, although this problem is nonconvex, embedding is very well studied and great amounts of effort have been made to design gradient-based training methods that can approximately find  a ``good'' local minimum. Here a ``good'' local minimum refers to a local minimum with a sufficiently low suboptimality gap.  Second, evaluating the embedding training objective and its gradient requires no calls to the imaging operator and is computationally inexpensive. This means that the supervised problem can be solved relatively quickly and easily when compared to Eq. \eqref{eqn:INR_obj}. 

\subsubsection{Continuous formulation} To leverage the existing capabilities for solving NF embedding problems, this work proposes a novel formulation of the spatiotemporal image reconstruction problem as a constrained minimization problem, in which the sought-after object estimate $f^*$ is constrained to the manifold $\mathcal{M}_{{\Phi}_{\boldsymbol \xi}} \coloneqq \{ f \in \mathbb{U} | f = {\Phi}_{\xi}, \  \xi \in \mathbb{R}^W\}$ of all square integrable spatiotemporal functions that can be represented exactly by the chosen NF architecture. That is, the object estimate $f^*$ is defined as
\begin{equation}
\label{eq:constrained_ip}
f^* \coloneqq \argmin_{f \in \mathcal{M}_{{\Phi}_{\boldsymbol \xi}}} \sum_{k=1}^{K} \mathcal{L}_k(f(\cdot, t_k) ) + \mathcal{R}(f),
\end{equation}
where, recalling Eq. \eqref{eq:continuous_data_fidelity}, the data fidelity term is given by
\begin{equation*}
\begin{split} 
\sum_{k=1}^{K} \mathcal{L}_k(f(\cdot, t_k) ) & = \sum_{k=1}^{K} \frac{1}{2\sigma^2} \left\| \mathcal{H}_k(f(\cdot, t_k) - \boldsymbol{d}_k\right\|^2 \\
{} & = \frac{1}{2\sigma^2} \left\| \mathcal{H}(f) - \boldsymbol{d}\right\|^2.
\end{split}
\end{equation*}

In this work, a stochastic generalized plug-and-play method, referred to as ProxNF, is developed to approximately solve Eqn. \eqref{eq:constrained_ip}, as summarized in Algorithm \ref{alg:Prox_Grad}. Plug-and-play (PnP) methods \cite{kamilov2017plug, sun2019online, xu2020provable, sun2021scalable}  are a generalization of the proximal gradient algorithm \cite{Rockafellar70,nitanda2014stochastic} in which the proximal operator is replaced by some denoising method.
To this aim, a generalized proximal operator (denoiser) that computes a regularized projection of a spacetime object estimate $f \in \mathbb{U}$ onto the manifold $\mathcal{M}_{{\Phi}_{\boldsymbol \xi}}$ is introduced. Its evaluation requires the solution of a regularized embedding problem defined as
\begin{equation}
\operatorname{Prox}_{\alpha\Reg}^{\mathcal{M}_{{\Phi}_{\boldsymbol \xi}}}(f) := \argmin_{\Tilde{f} \in \mathcal{M}_{{\Phi}_{\boldsymbol\xi}}} \frac{1}{2\alpha} \| \Tilde{f} - f\|_2^2 + \Reg(\tilde f).
\label{eq:proximal_projection}
\end{equation}

An update direction $g \in \mathbb{U}$ can then be computed, for example using gradient of the data-fidelity term, in the image domain  $\mathbb{U}$ and independent of the NF architecture. With the proximal operator, the updated estimate can then be projected back to the NF manifold and regularized. This can be summarized in the iterative process
\begin{equation}\label{eqn:prox_it_process}
    g^n = \nabla_f \LI(f^{n}), \ 
         f^{n+1} = \operatorname{Prox}_{\alpha^n\Reg}^{\mathcal{M}_{{\Phi}_{\boldsymbol \xi}}}(f^n - \alpha^n g^n),
\end{equation} where  $f^n$ represents the image estimate at the current iteration, $g^n$ the corresponding update direction at the $n$-th iteration, and $\alpha^n >0 $ is the (possibly) iteration dependent step-length.

This method then reframes the original self-supervised problem as a series of, easier to solve, supervised subproblems. In this work, the update direction $g$ is chosen to be the gradient of the data-fidelity term but could be generalized to include other choice of update direction (potentially conjugate gradient or (Quasi-)Newton directions). 

\subsubsection{Discretized formulation} Approximating the continuous-to-discrete imaging operator $\mathcal{H}(\cdot)$ with $\mathcal{H}_\Pi(\cdot) := \mathbf{H}(\Pi_{M,K} \cdot)$, the discretized counterpart of the minimization problem in Eqn. \eqref{eq:constrained_ip} reads
\begin{equation}
\label{eq:constrained_ip_discretized}
f_\Pi^* \coloneqq \argmin_{f \in \mathcal{M}_{{\Phi}_{\boldsymbol \xi}}} \sum_{k=1}^{K} \boldsymbol{L}_k(\Pi^{k}_{M,K}f ) + \mathcal{R}(f),
\end{equation}
where the data fidelity term is given by
\begin{equation*}
\begin{split}
\sum_{k=1}^{K} \boldsymbol{L}_k(\Pi^{k}_{M,K}f ) &= 
\sum_{k=1}^{K} \frac{1}{2\sigma^2} \left\| \mathbf{H}_k(\Pi^k_{M,K}f) - \boldsymbol{d}_k\right\|^2\\
{} & = \frac{1}{2\sigma^2} \left\| \mathbf{H}(\Pi_{M,K}f) - \boldsymbol{d}\right\|^2.
\end{split}
\end{equation*}
Above and in what follows, the subscript $\Pi$ in $f_\Pi^*$ is used to highlight the approximation introduced by replacing the continuous-to-discrete imaging operator $\mathcal{H}(\cdot)$ with the discretized one $\mathcal{H}_\Pi(\cdot)$.

The update direction $g_\Pi \in \mathbb{U}_{M,K}$, i.e. the counterpart of $g$ in Eq. \eqref{eqn:prox_it_process} for the discretized problem in Eq. \eqref{eq:constrained_ip_discretized}, is numerically computed as 

\begin{equation}
\begin{split}
g_\Pi^n & = \Pi_{M,K}^*\nabla_{\mathbf{F}}  \mathbf{L}(\Pi_{M,K} f^n) \\
{} & = \frac{1}{\sigma^2} \Pi_{M,K}^* \mathbf{H}^t \left( \mathbf{H}(\Pi_{M,K}f) - \boldsymbol{d} \right),
\end{split}
\label{eq:discretized_gradient_data_fidelity}
\end{equation}
where $\Pi_{M,K}^*: \R^{M\times K} \rightarrow \mathbb{L}^2(\Omega_T)$ is the adjoint of  $\Pi_{M,K}$ defined in Eqn \eqref{eqn:adjoint_projection}. However, explicitly forming the update direction $g_\Pi^n \in \mathbb{U}_{M,K}$ in Eqn. \eqref{eq:discretized_gradient_data_fidelity} requires to store an expansion coefficient matrix $\mathbf{G}^n := \nabla_{\mathbf{F}}  \mathbf{L}(\Pi_{M,K} f^n) \in \R^{M\times K}$ and thus can compromise the memory efficiency provided by the NF representation of the object. Memory efficiency can be preserved by computing an update direction $\tilde{g}_\Pi$ using stochastically subsampled frames. For a stochastically selected collection of frames $\{k_1,\hdots,k_J\}$ (with $J \ll K$), the update direction is then computed as 

\begin{equation}
\label{eq:discrete_stochastic_gradient}
\begin{split}
\tilde{g}_\Pi^n & = \Pi_{M,K}^* \frac{K}{J}\sum_{j=1}^J \nabla_{\mathbf{f}_{k_j}} \mathbf{L}_{k_j} (\Pi_{M,K}^{k_j} f^n)\otimes \mathbf{e}_{k_j} \\
 &= \frac{K}{J}\sum_{j=1}^J \left( \Pi_{M,K}^{k_j}\right)^* \nabla_{\mathbf{f}_{k_j}} \mathbf{L}_{k_j} (\Pi_{M,K}^{k_j} f^n) \\
{} & = \frac{K}{J} \sum_{j=1}^J \left( \Pi_{M,K}^{k_j}\right)^* \boldsymbol{g}_{k_j}^n, 
\end{split}
\end{equation}
with
$$\boldsymbol{g}_{k_j}^n = \nabla_{\mathbf{f}_{k_j}} \mathbf{L}_{k_j} (\Pi_{M,K}^{k_j} f^n) = \mathbf{H}_{k_j}^t \left( \mathbf{H}_{k_j} \Pi_{M,K}^{k_j} f^n - \boldsymbol{d}_k \right). $$

Representing $\tilde{g}_\Pi^n$ only requires storing an expansion coefficient matrix $\tilde{\mathbf{G}}^n \in \R^{M\times J}$, whose $j$-th column is $\boldsymbol{g}_{k_j}^n$.

This proposed method is summarized in Algorithm \ref{alg:Prox_Grad}, where the subscript $\Pi$ is omitted for ease of notation. 
Note that in the algorithm the spatiotemporal function $f^n$ ($n \geq 0$) is represented using an NF and only the network parameters $\boldsymbol{\xi}^n \in \mathbb{R}^{N_p}$ need to be stored. Similarly, the spatiotemporal function $\tilde{g}^n$ ($n \geq 0$) stemming from the randomly subsampled gradient of the data fidelity term is represented by the coefficients of the dense matrix $\tilde{\mathbf{G}}^n \in \mathbb{R}^{M\times J}$ with $J \ll K$. Thus, Algorithm \ref{alg:Prox_Grad} can be implemented in a memory-efficient manner, without the need to store an explicit representation of the spatiotemporal object estimate as an $\mathbb{R}^{M\times K}$ coefficient matrix.

\begin{rem}
Algorithm \ref{alg:Prox_Grad} finds an approximate solution to the minimization problem in Eqn. \eqref{eq:constrained_ip_discretized} by use of proximal gradient descent (i.e., the so-called forward-backward splitting). In principle, also the alternating direction method of multipliers (ADMM, \cite{boyd2011distributed}) or backward-backward splitting (e.g. Douglas-Rachford algorithm, \cite{parikh2014proximal,lions1979splitting}) could be employed to find an approximate solution of Eqn. \eqref{eq:constrained_ip_discretized}. However, it is worth to note that while Algorithm \ref{alg:Prox_Grad} allows for a memory-efficient implementation that only requires storing the NF parameters $\boldsymbol{\xi}$ and the vectors $\boldsymbol{g}^n_{k_j}$, this is not possible using ADMM. In fact, updating the Lagrangian multiplier in the ADMM method would require the solution of an extra embedding step.
\end{rem}

\begin{algorithm2e}\label{alg:Prox_Grad}
\SetAlgoLined

Initialize neural field $f^0 = \Phi_{\boldsymbol \xi}$. \\

\For{$n=0,1,2,\hdots$}{
Randomly select collection of frames $\{k_1,\hdots, k_J\}.$
\For{$j=1,\hdots, J$}{
Render the NF at the $k_j$-th frame

$$\mathbf{f}_{k_j}^n = \Pi_{M,K}^{k_j} f^n$$

Compute the gradient of the discretized data fidelity term w.r.t. the $k_j$-th frame coefficients

$$\mathbf{g}_{k_j}^n = \nabla_{\mathbf{f}_{k_j}} \mathbf{L}_{k_j}(\mathbf{f}_{k_j}^n).$$ }
Form the update direction according to Eqn. \eqref{eq:discrete_stochastic_gradient}

$$\tilde{g}^n = \frac{K}{J}\sum_{j=1}^J  \left(\Pi_{M,K}^{k_j}\right)^*\mathbf{g}_{k_j}^n.$$

Adaptively select the step size $\alpha^n$.

$$\alpha^n = \sigma^2 \frac{ \|\tilde{g}^n \|^2}{\sum_{j=1}^J\|\mathbf{H}_{k_j} \Pi_{M,K}^{k_j}\tilde{g}^n\|^2 }.$$

Apply proximal update in Eqn. \eqref{eq:proximal_projection}

\begin{equation}\label{eq:ProxNF_iterate}
    f^{n+1} \leftarrow \operatorname{Prox}^{\mathcal{M}_{\Phi_{\boldsymbol{\xi}}}}_{\alpha^n\Reg}(f^n - \alpha^n \tilde{g}^n)
\end{equation}

}

 \caption{ProxNF: Spatiotemporal image reconstruction with neural fields via proximal splitting}
\end{algorithm2e}

\subsection{Convergence of ProxNF}

This section provides an insight on the convergence properties of Algorithm \ref{alg:Prox_Grad} in the general framework of online PnP algorithms\cite{KamilovPNPreviewPaper} by use of well-established mathematical tools from traditional stochastic optimization and monotone operator theory. 

Consider the minimization problem
\begin{equation}\label{eq:cost_conv_analyis}
f^* = \operatornamewithlimits{argmin}_{f\in \mathcal{M}}{\sum_{k=1}^K \mathcal{L}_k(f) + \mathcal{R}(f)},
\end{equation}
where
\begin{itemize}
\item[A.1] The data fidelity terms $\mathcal{L}_k(\cdot)$ are convex, differentiable functions with Lipschitz continuous gradients and Lipschitz constant $C > 0$;
\item[A.2] The regularization term $\mathcal{R}$  is a closed proper function;
\item[A.3] The set $\mathcal{M}\subset \mathbb{U}$ is convex.
\end{itemize}

Define the operator
\begin{equation}
P(f) \coloneqq {\rm prox}^\mathcal{M}_{\alpha\mathcal{R}}\left(f - \alpha \frac{K}{J}\sum_{j=1}^J \nabla_f \mathcal{L}_{k_j}(f)\right),
\end{equation}
where the operator ${\rm prox}_{\alpha\mathcal{R}}^{\mathcal{M}}$ is defined analogously to Eq.\eqref{eq:proximal_projection} and $\{k_1, \ldots, k_J\}$ (with $J\ll K$) is a collection of randomly selected frames.

Under the assumptions above, Proposition \ref{prop:converge} establishes the convergence of Algorithm \ref{alg:Prox_Grad} to a fixed point of the operator $P$, when the non-convex manifold $\mathcal{M}_{\Phi_{\boldsymbol{\xi}}}$ is replaced by the convex set $\mathcal{M}$.

\begin{prop}
\label{prop:converge}
Under Assumptions A.1-A.3 and with a proper choice of the step size $\alpha \in (0, 1/C]$, the iteration $f^{n+1} = P(f^n)$ converges in expectation to a fixed point of $P$, and thus to a minimizer of Eq. \eqref{eq:cost_conv_analyis}, at worse sublinearly.
\end{prop}

\begin{proof}
Under assumptions [A.2] and [A.3], the operator ${\rm prox}_{\alpha\mathcal{R}}^{\mathcal{M}}$ is \emph{firmly non-expansive} \cite{parikh2014proximal}, and thus $\theta$-averaged with $\theta=1/2$. Furthermore, the gradient estimate $\tilde{g} = \frac{K}{J}\sum_{j=1}^J \nabla_f \mathcal{L}_{k_j}(f)$ is unbiased and has bounded variance. Therefore, all the conditions of Assumption 2 in \cite{sun2019online} hold. The convergence of the fixed-point iteration $f^{n+1} = P(f^n)$ then follows from Proposition 5 and Corollary 1 in \cite{sun2019online}. The equivalence between the set of the fixed points of $P$ and the minimizers of Eq. \eqref{eq:cost_conv_analyis} follows from A.1--A.3 and the definition of the proximal operator \cite{sun2019online}.
\end{proof}

\begin{rem}
\label{rem:convergence}
Proposition \ref{prop:converge}  provides theoretical convergence guarantees of Algorithm \ref{alg:Prox_Grad} when the NF manifold $\mathcal{M}_{\Phi_{\boldsymbol{\xi}}}$ is a convex set. While not true for a general NF architecture, this holds for the POUnet architecture in Eq. \eqref{eq:pounet_def} with frozen partitioning parameters $\boldsymbol{\eta}$. Therefore, convergence to a fixed point can be theoretically guaranteed by modifying the proximal operator in Eq. \eqref{eq:ProxNF_iterate}, to only update the coefficients $\boldsymbol{C}$ keeping $\boldsymbol{\eta}$ fixed.  Empirical evidence of the convergence in the general case is provided in the numerical results, where it was observed that also the coefficients $\boldsymbol{\eta}$ quickly stabilize after a few ProxNF iterations.
\end{rem}

\begin{rem}
Nesterov acceleration is not used Algorithm \ref{alg:Prox_Grad}. This is because proximal splitting methods combining randomized evaluation of the data-fidelity gradient and/or ordered subsets with Nesterov acceleration may be susceptible to instabilities due to accumulation of errors in the momentum term {\cite{kim2013ordered, kim2014combining, haase2020improved, sisniega2021accelerated}}.  
\end{rem}

\subsection{Adapted partition of unity neural fields}

This work also introduces a modified partition of unity network of the form

\begin{equation}
\label{eq:mpounet_def}
\Phi_{(\boldsymbol \eta,\boldsymbol C)}(\rr,t) = \boldsymbol{\Psi}^T_{\boldsymbol{\eta}}(t)^T \boldsymbol{C}^T \boldsymbol{B}(\rr),
\end{equation}
where the partition of unity network $\boldsymbol{\Psi}_{\boldsymbol{\eta}}$ only depends on the time coordinate $t$ and the basis functions in $\boldsymbol{B}$ only depends on the spatial coordinate $\rr$. In particular, $\boldsymbol{B}(\rr) = \begin{pmatrix}
    \beta_1(\rr)  & \hdots & \beta_M(\rr)
\end{pmatrix}$ shares the same basis functions $\{\beta_m\}_{m=1}^M$ in Eqn. \eqref{eqn:disc_decomp}.
This representation allows for the high spatial resolution of conventional discretization methods and smooth temporal transitions while  exploiting spatiotemporal redundancies. Moreover, this architecture greatly simplifies discrete projection operations. That is for a selected frame $k$, $$\Pi_{M,K}^k \Phi_{(\boldsymbol{\eta},\boldsymbol{C})}=  \mathbf{C} \boldsymbol{\Psi}_{\boldsymbol{\eta}}(t_k).$$

In this work, the regularization term $\Reg$ was chosen to be Tikhonov regularization  on the spatial derivatives computed with finite difference and on the temporal derivatives computed with automatic differentiation of the learned partition of unity.  This choice of regularization then allowed the proximal $\operatorname{Prox}_{\alpha,\Reg}^{\textnormal{NF}}$ to be computed using the alternating minimization algorithm provided in Eq. \eqref{eqn:alt_min}, where the coefficient updates can be accomplished by solving a linear system while the partition can be updated using a stochastic gradient method.

\section{Numerical Studies}\label{sec:numerical_studies}

This work utilized two numerical studies to assess the proposed NF representation and proposed ProxNF method for spatiotemporal image reconstruction. These studies were inspired by dynamic contrast-enhanced photoacoustic computed tomography (DCE PACT) imaging of tumor perfusion in small animal models \cite{hupple2018light}. PACT is a hybrid biomedical imaging modality that combines benefits from both optical and ultrasound imaging\cite{WangYao16}. Specifically, PACT combines the benefits of high soft tissue contrast from optical imaging and high-resolution detection principles from ultrasound imaging to create high-resolution maps of optical properties. PACT image formation utilizes a short laser pulse in the near infrared range to illuminate a biological tissue. The biological tissue then absorbs the optical energy and generates heat and a local increase in pressure, which is described as the photoacoustic effect. This pressure distribution then propagates as an acoustic wave where it can be measured outside the tissue using ultrasound detectors. Under the assumption of a non-lossy homogeneous acoustic medium, that is, a medium with homogeneous speed of sound and density and no acoustic attenuation, the relationship between the induced pressure distribution and the data can be modeled using the circular Radon transform in two spatial dimensions and the spherical Radon transform in three spatial dimensions \cite{MatthewsAnastasio2017, WangAnastasio15}. The numerical studies presented in this work, focus on estimation of this spatially and temporally varying induced pressure distribution from sparse-view measurements collected from a circular aperture in two spatial dimensions and a cylindrical aperture in three spatial dimensions. 

\subsection{Construction of the dynamic object}\label{subsec:dynamic_object} 
\begin{figure}
    \centering
    \includegraphics[width = 0.9\columnwidth, trim = {4cm 0cm 3cm 1cm}, clip]{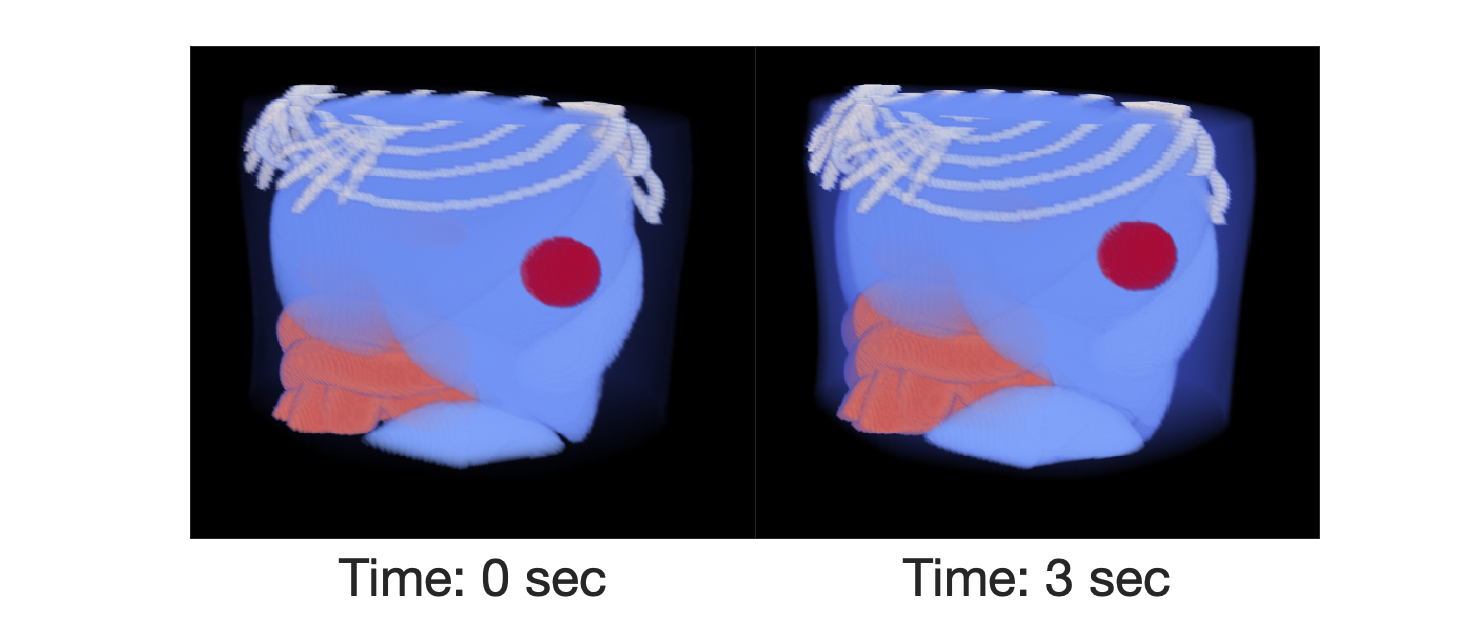}
    \caption{Numerical dynamic mouse abdominal phantom extract from mouse whole body phantom (MOBY) and shown at the beginning and end of inspiration during a 5 second breathing period. The phantom is comprised of  detailed three-dimensional (3D) anatomical structures featuring several organs and models physiologically realistic cardiac and respiratory motions. A spherical lesion of diameter 7 mm was inserted in the abdominal region at a depth of 5 mm to mimic a subcutaneous tumor flank and displayed in dark red. }
    \label{fig:anat}
    \vspace{-5mm}
\end{figure}

The dynamic objects used in these studies were spatiotemporal maps of the spatially and temporally varying induced pressure distribution within a numerical tumor perfusion model in two and three spatial dimensions. The anatomical structures of these objects were constructed utilizing an abdominal section extracted from an anatomically realistic whole-body murine numerical phantom (MOBY) \cite{SegarsTsuiFreyEtAl04,SegarsTsui2009}. MOBY comprises detailed three-dimensional (3D) anatomical structures featuring tens of organs and models physiologically realistic cardiac and respiratory motions and is illustrated in Figure \ref{fig:anat}. A spherical lesion of diameter 7 mm was inserted in the abdominal region at a depth of 5 mm to mimic a subcutaneous tumor flank. To simulate DCE imaging of a tumor perfusion model, the concentration of an optical contrast agent in the lesion and blood-filled organs was computed using a two compartment Toft's model \cite{tofts1997modeling, tofts1999estimating, sourbron2011scope}. In this model, the contrast agent was introduced after 90 seconds and injected over a 30 second window following a slow bolus pattern.  Optical properties (optical absorption coefficient and scattering coefficient) corresponding to an 800 nm laser pulse wavelength were then assigned to each tissue type based on physiologically plausible chromophore concentrations (oxygenated hemoglobin, deoxygenated hemoglobin, water, fat, contrast agent)\cite{jacques2013optical}. Contrast agent optical properties were chosen to mimic those of indocyanine green (ICG) \cite{alander2012review}, in which case the absorption spectrum has a peak near the chosen wavelength.

The diffusion approximation\cite{tarvainen2012reconstructing} was used to simulate photon transport  at each imaging frame. The incident fluence was assumed constant on the surface of the object. The fluence distribution within the object was obtained by solving the partial differential equation stemming from the diffusion approximation using the open-source finite element library FEniCS \cite{LoggMardalWells122}, thus yielding a volumetric map of dynamic induced pressure distribution at each imaging frame. For the experiments in two spatial dimensions, a slice was extracted from the 3D induced pressure phantom.

\subsection{Definition of the virtual imaging systems}
\label{subsec:imaging_system}

\subsubsection{Dynamic imaging system for two spatial dimensions}
\label{subsec:imaging_system_2d}

\begin{figure}
    \centering
    \includegraphics[width= \columnwidth]{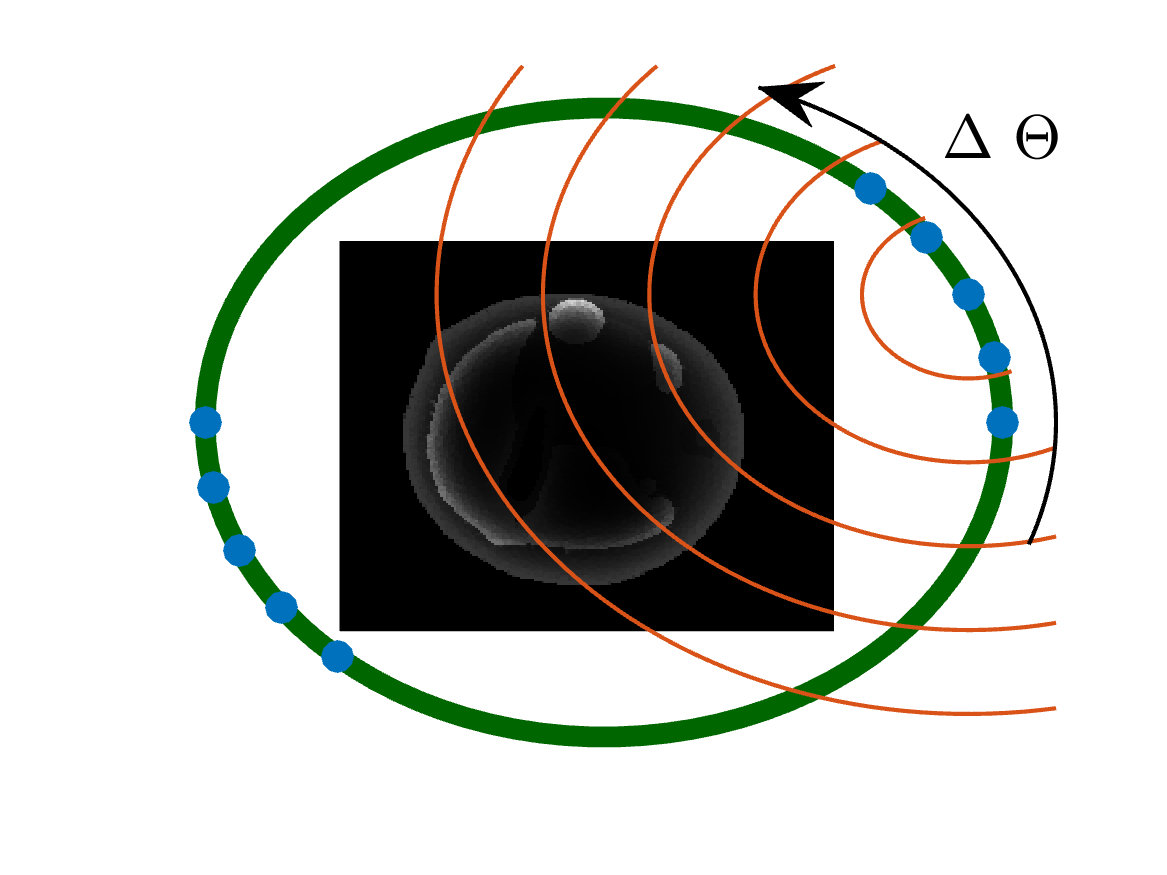}
    \vspace{-15mm}
    \caption{Illustration of the virtual system for dynamic imaging using a CRT model. At each imaging frame, data are collected by idealized point-like transducers (shown in blue). Transducers rotate an angle of $\Delta \Theta$ between frames.}
    \label{fig:2dsystem}
    \vspace{-5mm}
\end{figure}

Figure \ref{fig:2dsystem} illustrates the virtual imaging system used in the numerical study with two spatial dimensions. The measurement geometry was a circle of radius $R$ (shown in green), along which two groups of 5 idealized point-like ultrasound transducers are distributed. The sensors in each group were spaced $1^\circ$ apart on the transducer arc. Each group of sensors was distributed antipodal from one another. The object was placed at the center of the imaging system and fully contained within a square region of size $L$ (field of view). At each imaging frame, data were collected simultaneously by each sensor corresponding to the integral along $I$ concentric arcs. Sensors were rotated by an angle $\Delta \Theta$ after each imaging frame. The dynamic changes of the object were assumed to be slow enough such that the object was not expected to vary significantly during the acquisition of each imaging frame (quasi-static approximation). This assumption is often valid for several dynamic PACT applications since the duration of the light excitation pulse at each imaging frame is in the range of a few nanoseconds while the dynamic changes in the object have a time scale in the range of milliseconds to seconds.  

Under these assumptions, the imaging operator in two spatial dimensions can then be described by the circular Radon transform (CRT).  
The discrete-to-discrete counterpart of the CRT imaging operator was implemented using the AirToolsII MatLab toolbox \cite{HansenJorgensen2018} and stored as a compressed sparse column matrix\cite{gilbert1992sparse}. The imaging systems parameters are displayed in Table \ref{Tab:is_parameters_comb}.  

\subsubsection{Dynamic imaging system for three spatial dimensions}
\label{subsec:imaging_system_3d}

For the numerical study in three spatial dimensions, the measurement geometry was a cylinder of radius $R$ and height $A$, along which four groups of idealized point-like ultrasound transducers were distributed. Each group of sensors consisted of 5 columns of sensors spaced $1^\circ$ apart. Each column consisted of 96 sensors spaced $\Delta z$ apart. These four groups of sensors were evenly spaced $90^\circ$ apart.

The object was placed at the center of the imaging system and fully contained within a rectangular prism region of side length $L$ in the $x$ and $y$ direction and height $H$. At each imaging frame, data were collected at each sensor ($I$ measurements per sensor). Sensors rotated an angle $\Delta \Theta$ along the z-axis of the cylindrical geometry at each frame. Again, the dynamic changes in the object were assumed to be slow enough such that the object is approximately constant at each imaging frame.

Under these assumptions, the imaging operator in three spatial dimensions can be described by the spherical Radon transform (SRT).  The SRT, which computes the integral of the object function over concentric spherical shells, is the 3D generalization of the CRT. To construct the discrete-to-discrete counterpart of the  SRT operator, this work utilizes a computationally efficient method, described in \cite{Lozenski2024cylSRT}, which exploits the symmetry of the cylindrical measurement geometry and decomposes the SRT as the composition of two 2D CRT operators \cite{HaltmeierSRT17}. The imaging system parameters for these experiments are reported in Table \ref{Tab:is_parameters_comb}.

\begin{table}
\vspace{0mm}
\centering
\caption{Dynamic imaging system parameters} 
\label{Tab:is_parameters_comb}
\begin{tabular}{ll}
\hline
\multicolumn{2}{c}{Imaging system}\\
FOV diameter $L$ & 3.72 cm\\
FOV height $H$ (3D) & 2 cm \\ 
Aperture $R$ & 2.63 cm \\
Aperture height $A$ & 4 cm \\
Acquisition time $T$ &  648 s \\ 
Number of frames $K$ & 1,296 \\ 
Sensor separation & $1^\circ$ \\
Rotation angle $\Delta\Theta$ & $5^\circ$ \\
Number of sensors per group  & $5$\\ 
\hline
\multicolumn{2}{c}{Two spatial dimensions}\\
Number of pixels $M = M_s^2$ & $186^2$ \\ 
Rings per view $I$ & 263 \\
Relative noise & $0.04$\\
Number of sensor groups & 2\\
Number of views per frame $S$ & $10$ \\
Total number of sensors & 10\\
\hline
\multicolumn{2}{c}{Three spatial dimensions}\\
Number of voxels $M = M_s^2M_z$ & $186^2 \times 100$ \\ 
Rings per view $I$ & 323 \\
Relative noise & $0.01$\\
Number of sensor groups & 4\\
Number of views per frame $S$ & $20$ \\
Number of sensors per column $Z$ & 96 \\ 
Total number of sensors & 1,920\\
\hline
\end{tabular}
\vspace{-4mm}
\end{table}

\subsection{Study Design}

\paragraph*{Study 1: Dynamic embedding}

The first numerical study solved the embedding problem and assessed the proposed NF architecture for representing the dynamic object described in Section \ref{subsec:dynamic_object}. 
For the case involving two spatial dimensions, the proposed NF architecture was compared to the original NF architecture with polynomial basis functions \cite{LozenskiNeuralFields} and a low rank (semiseparable) decomposition of the dynamic object. The proposed POUnet architecture used a  network $\boldsymbol{\Psi}_{\boldsymbol{\eta}}$ with 4 hidden layers of width 140 nodes and one output layer of width $P=10$ partitions. Sinusoidal and softmax activation functions were employed in the hidden and output layers, respectively. The set of basis function $\mathbf{B}$ consisted of $L = 186^2 = 34596$ pixel indicator functions. This resulted in a representation with $427,020$ parameters. The original POUnet architecture used a network $\boldsymbol{\Psi}_{\boldsymbol{\eta}}$  with 5 hidden layers of width 280 nodes and an output layer of width $P=100$ partitions. Third order space-time separable polynomials were employed as basis functions ($L=40$). This resulted in a representation with $425,800$ parameters. These two NF representations were compared to a semiseparable approximation of rank $r = 12$, SS(12), which has $430,704$ parameters. These representations allow for a 40-fold reduction in the number of  parameters compared to representing the object as a collection of spatiotemporal voxel values ($\sim 20$ million parameters).

For the case involving three spatial dimensions, the proposed NF architecture was compared to a low rank, semiseparable, decomposition of the dynamic image.  The POUnet architecture again utilized the same  network $\boldsymbol{\Psi}_{\boldsymbol{\eta}}$ as the 2D case and a set of basis function $\mathbf{B}$ that contained $L=186^2\cdot100 = 3,459,600$ voxel indicator functions. This resulted in a representation with $34,988,424$ parameters. The semiseparable approximation used a rank $r = 10$, SS(10), which has $34,608,960$ parameters. 
These representations yield over two orders of magnitude reductions in the number of parameters compared to representing the object as a collection of spatiotemporal voxel values ($\sim 5$ billion parameters).

\paragraph*{Study 2: Dynamic reconstruction}

 The second numerical study explored dynamic image reconstruction in two (CRT imaging operator) and three (SRT imaging operator) spatial dimensions. In both cases, two spatiotemporal image reconstruction approaches were utilized. The first reconstruction technique utilized the proposed NF architecture and the proposed ProxNF method in Algorithm \ref{alg:Prox_Grad}. The second reconstruction technique, which serves as a reference, was spatiotemporal image reconstruction with nuclear norm regularization (STIR-NN) \cite{WangXiaetal14}. In particular, for the case involving three space dimensions, the memory-efficient maximal-rank-constrained method in \cite{cam2024spatiotemporal} was implemented. For the two spatial dimensions case, the NF approach in \cite{LozenskiNeuralFields}, which directly minimizes the objective in Eq. \eqref{eqn:INR_obj} using gradient descent, was also considered. 
 
 Measurement data, virtually acquired as described in Sect. \ref{subsec:imaging_system}, were corrupted with \emph{i.i.d.} additive Gaussian noise with zero mean and standard deviation $\sigma$ controlled by a specified relative noise level (RNL), defined as $\operatorname{RNL} = \sigma/\|\data\|_\infty$. In the two spatial dimensions case, the RNL was set to 0.04 and, in the three spatial dimensions case, the RNL was set to 0.01. Regularization weights for all reconstructions were chosen using Morozov's discrepancy principle \cite{Morozov66}, which selects the regularization weight such that the optimal data-fidelity value is approximately equal to the noise standard deviation.

\paragraph*{Evaluation metrics} In these studies, image quality  was quantified using relative root mean square error (RRMSE) and structural similarity index measure (SSIM) \cite{WangBoviketall04}, evaluated both on the entire image and on a region of interest (ROI) containing the lesion. Image quality was also assessed by analyzing the estimated lesion activity curve (LAC), which is the time-varying spatially-averaged signal over a volume of interest (the lesion). 

\section{Results}\label{sec:results}

The numerical studies in two spatial dimensions were performed on a workstation with two Intel Xeon Gold 5218 Processors (16 cores, 32 threads, 2.3 GHz, 22 MB cache each), 384 GB of DDR4 2933Mhz memory, and one NVidia Titan RTX 24GB graphic processing unit (GPU).  The numerical studies in three spatial dimensions were performed on the \emph{Delta} supercomputer at the National Center for Supercomputing Applications (NCSA) of the University of Illinois at Urbana-Champaign. Specifically, a single node with dual AMD Milan processors (64 core, 128 threads, 2.55 GHz, 96MB cache each), 256 GB of DDR4 3200 MHZ memory and two quad A100 40GB GPUs was used in the studies below.

The proposed method, ProxNF, was implemented using PyTorch, an open-source machine learning framework \cite{NEURIPS2019_9015} and executed on GPUs. The POUnet parameters $\boldsymbol{\eta}$ and $\boldsymbol{C}$ were initialized by performing an embedding of the time-averaged object for the representation studies and by solving a static image reconstruction problem on time-averaged measurements for the reconstruction studies. 

Following initialization, the NFs in the representation study were trained using the alternating minimization approach in Eq. \eqref{eqn:alt_min}.  Similarly, after initialization, the NFs in the reconstruction study were trained with the proposed proximal splitting method in Algorithm \ref{alg:Prox_Grad}. A reduction of two orders of magnitude in the norm of the proximal gradient was used as a stopping criterion. The batch size was $J=324$ and time frames were stochastically selected at each iteration.  Tikhonov regularization was used to penalize the $L^2$ norm of the gradient of the spatiotemporal image. The regularized embedding problem at each proximal update was solved as follows. The linear systems corresponding to the updates of $\boldsymbol{C}$ were solved using the  conjugate gradient method. The internal updates for $\boldsymbol{\eta}$ were accomplished using an Adam optimizer \cite{KingmaBa2014}. Specifically, 10,000 iterations with a fixed learning rate of $10^{-5}$ and batch size of 100,000  randomly sampled spatiotemporal points were used.

In the two-dimensional case, an additional NF was trained utilizing the method in \cite{LozenskiNeuralFields}, referred in what follows as GradNF. GradNF directly minimizes the objective in \eqref{eqn:INR_obj} using gradient descent and the alternating minimization algorithm in Eq. \eqref{eqn:alt_min}. 
For ease of comparison, GradNF utilized the same network architecture and randomization approach as the ProxNF method. The internal updates of $\boldsymbol{C}$ and $\boldsymbol{\eta}$ were similarly accomplished using an Adam optimizer \cite{KingmaBa2014} with a fixed learning rate of $10^{-5}$ and batch sizes of $1.1.\times 10^{7}$ corresponding to spatiotemporal voxels for each imaging frame. A reduction of two orders of magnitude in the norm of the gradient was used as a stopping criterion.

The spatiotemporal reconstructions in two spatial dimensions using the reference method (STIR-NN) were accomplished utilizing  the Fast Iterative Shrinkage Algorithm (FISTA) \cite{BeckTeboulle09} as implemented in UNLocBoX, a MATLAB toolbox for proximal-splitting methods \cite{PerraudinKalofoliasShuman2014}. The low rank semiseparable approximation of the image with two spatial dimensions was accomplished using the SVD command in MATLAB and soft-thresholding. 
The spatiotemporal reconstructions in three spatial dimensions using the reference method (STIR-NN) were accomplished using the memory-efficient randomized algorithm in \cite{cam2024spatiotemporal} with a maximum rank constraint of 32.
The randomized singular value decomposition \cite{HalkoMartinssonTropp11} was employed to compute the low rank decomposition and enforce the maximum rank constraint.

\subsection{Study 1: Dynamic embedding problem}

In this study, the NF representations  and semiseparable approximations of  rank $r$, SS($r$), were compared for a 2D cross-sectional slice and 3D volume extracted from the dynamic tumor perfusion phantom described in Section \ref{subsec:dynamic_object}.

\begin{figure}
    \centering
    \includegraphics[width =  0.8\columnwidth]{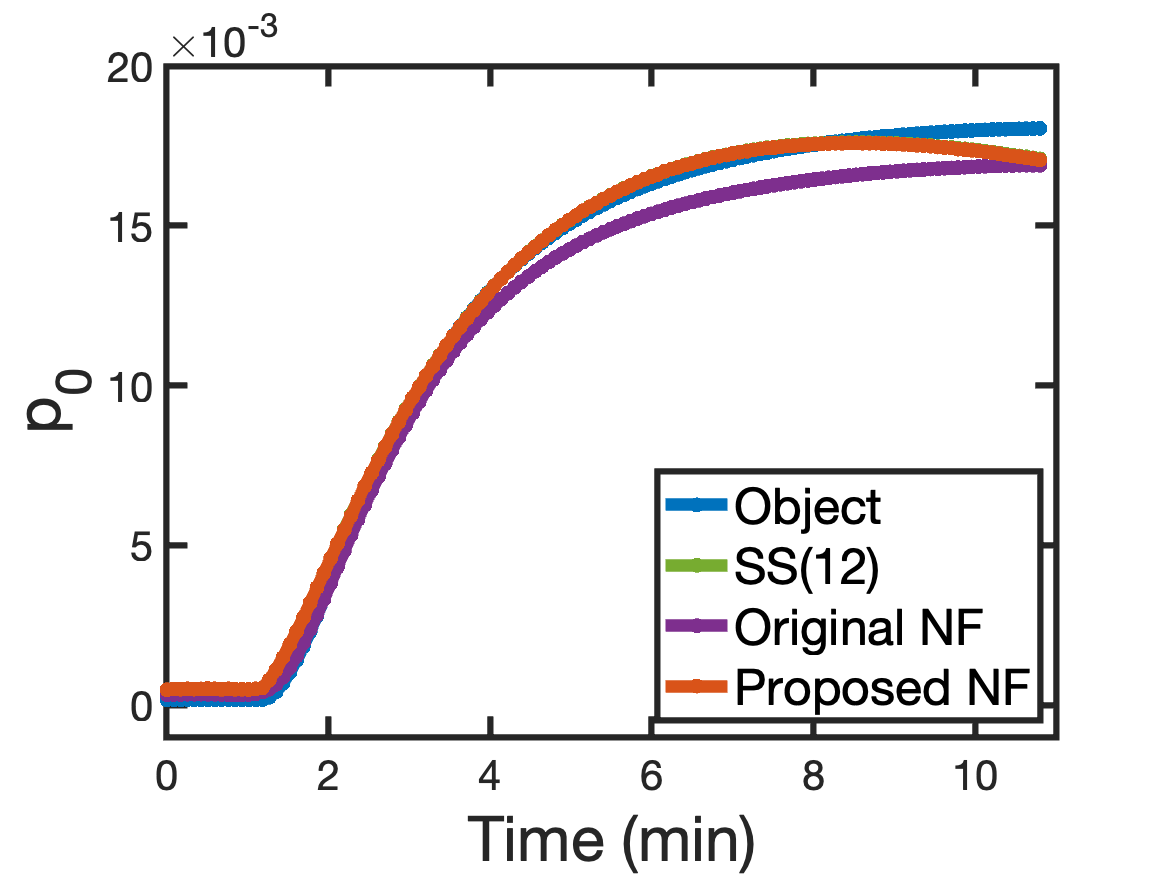}
    \caption{Study 1a: Dynamic embedding in two spatial dimensions. Lesion activity curve (LAC) of the object (blue), SS(12) approximation (green), original POUnet architecture (purple), and proposed POUnet architecture (red). The proposed POUnet architecture  better captures the dynamic changes of the object compared to the original POUnet architecture in \cite{LozenskiNeuralFields}. }
    \label{fig:rep_activity}
    \vspace{-5mm}
\end{figure}

\begin{figure*}
    \centering

    \includegraphics[width = 0.8\textwidth, trim = {1cm 0cm 4cm 1cm},clip]{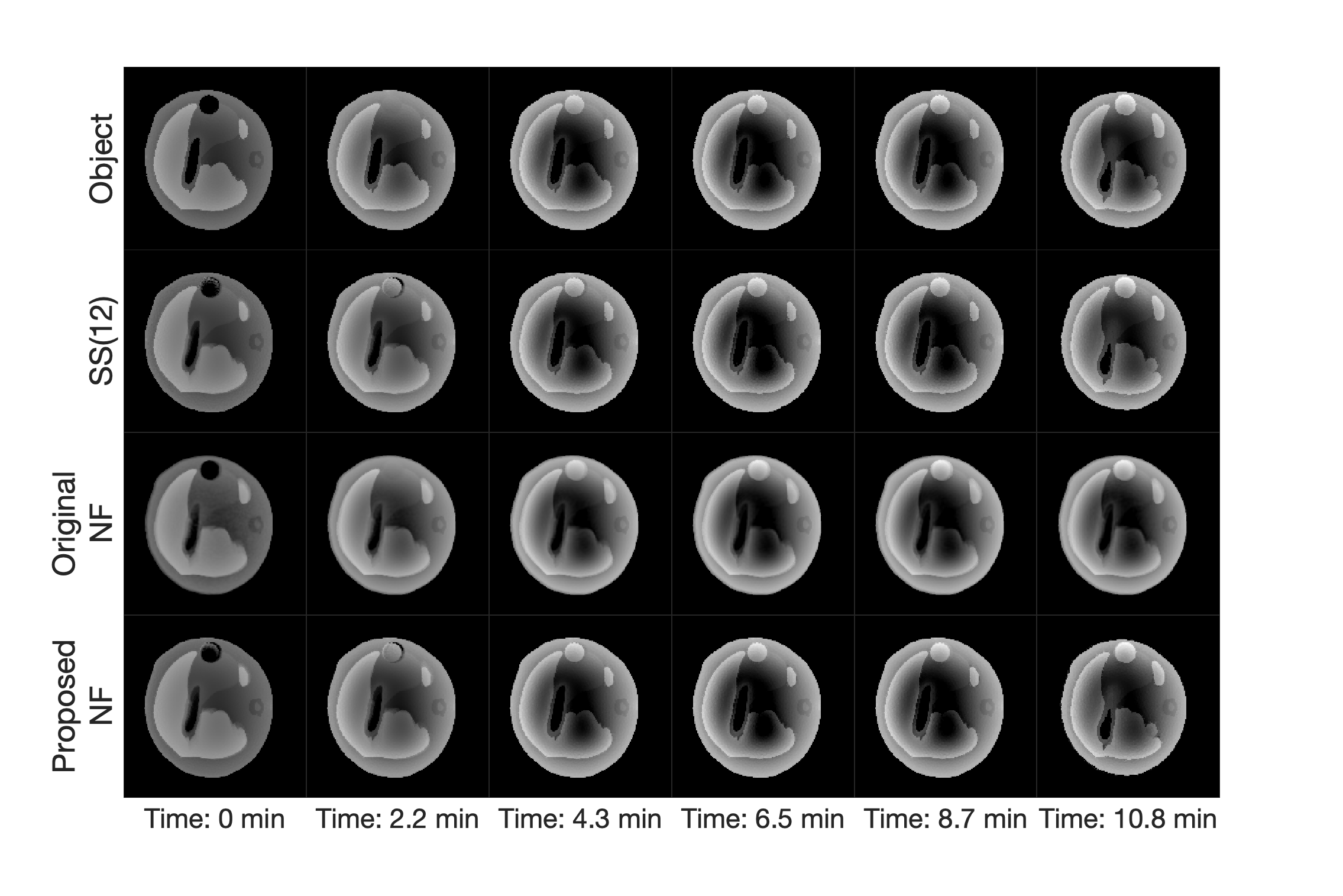}
    \vspace{-10mm}
    \caption{Study 1a: Dynamic embedding in two spatial dimensions. 
    Object (top), Rank 12 semiseparable approximation SS($12$) (second row),  NF using original POUnet architecture (third row), and NF using proposed POUnet architecture are shown at selected frames throughout the time period. The dynamic range of the grayscale image is $[10^-4, 0.06]$. A logarithmic scale was used to increase visualization contrast at depth. An animation of each representation constructed in this study is available in the supplemental multimedia materials (Video 1).}
    \label{fig:2DRepresentation}
\end{figure*}

\begin{figure}
    \centering
    \vspace{-0.5cm}
    \includegraphics[width = 0.8\columnwidth]{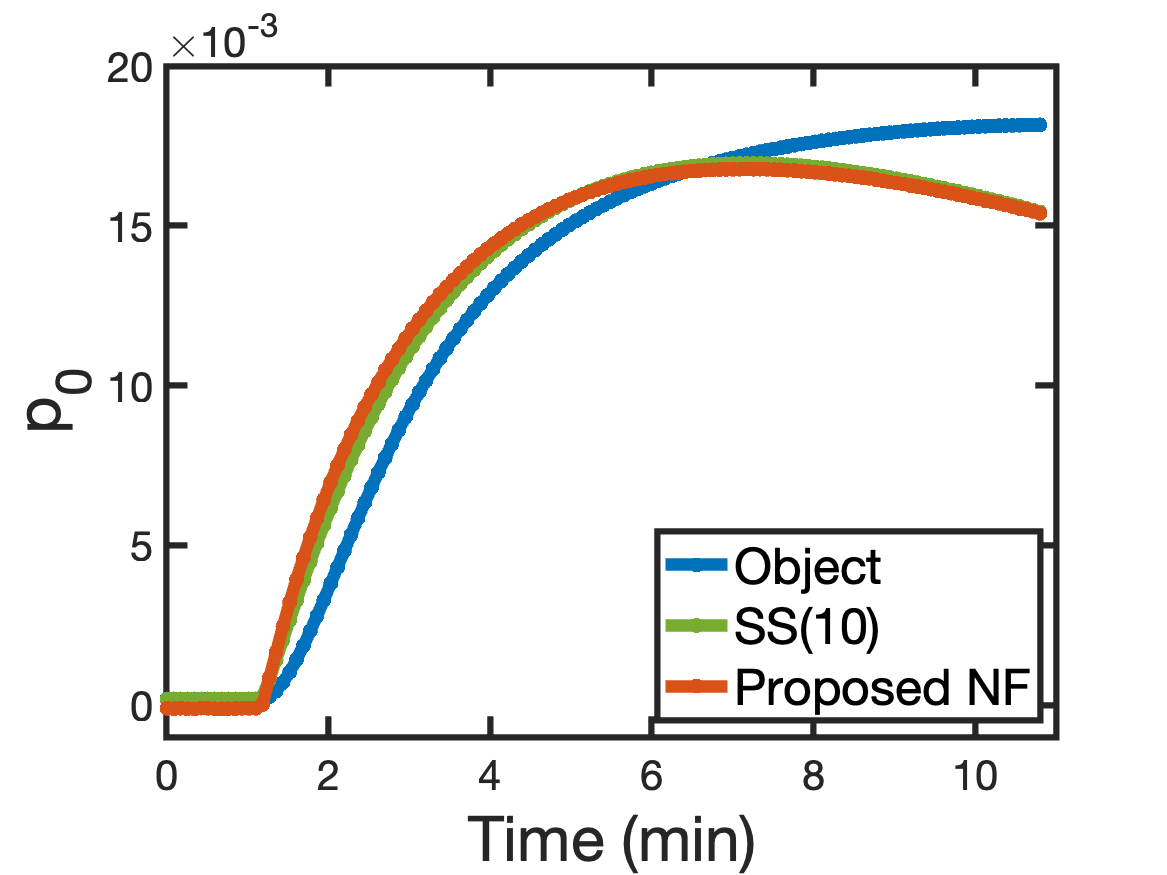}
    \caption{Study 1b: Dynamic embedding in three spatial dimensions. Lesion activity curve (LAC) of the object (blue), the semi-separable rank 10 approximation (green), and the NF representation (red). The semiseparable and NF representations lead to  almost identical, smooth LACs that underestimates activity at later times.}
    \label{fig:3dRepAct}
    \vspace{-5mm}
\end{figure}

\begin{figure*}
    \centering
    \includegraphics[width = 0.8\textwidth, trim = {2.5cm 0cm 2.5cm 0cm},clip]{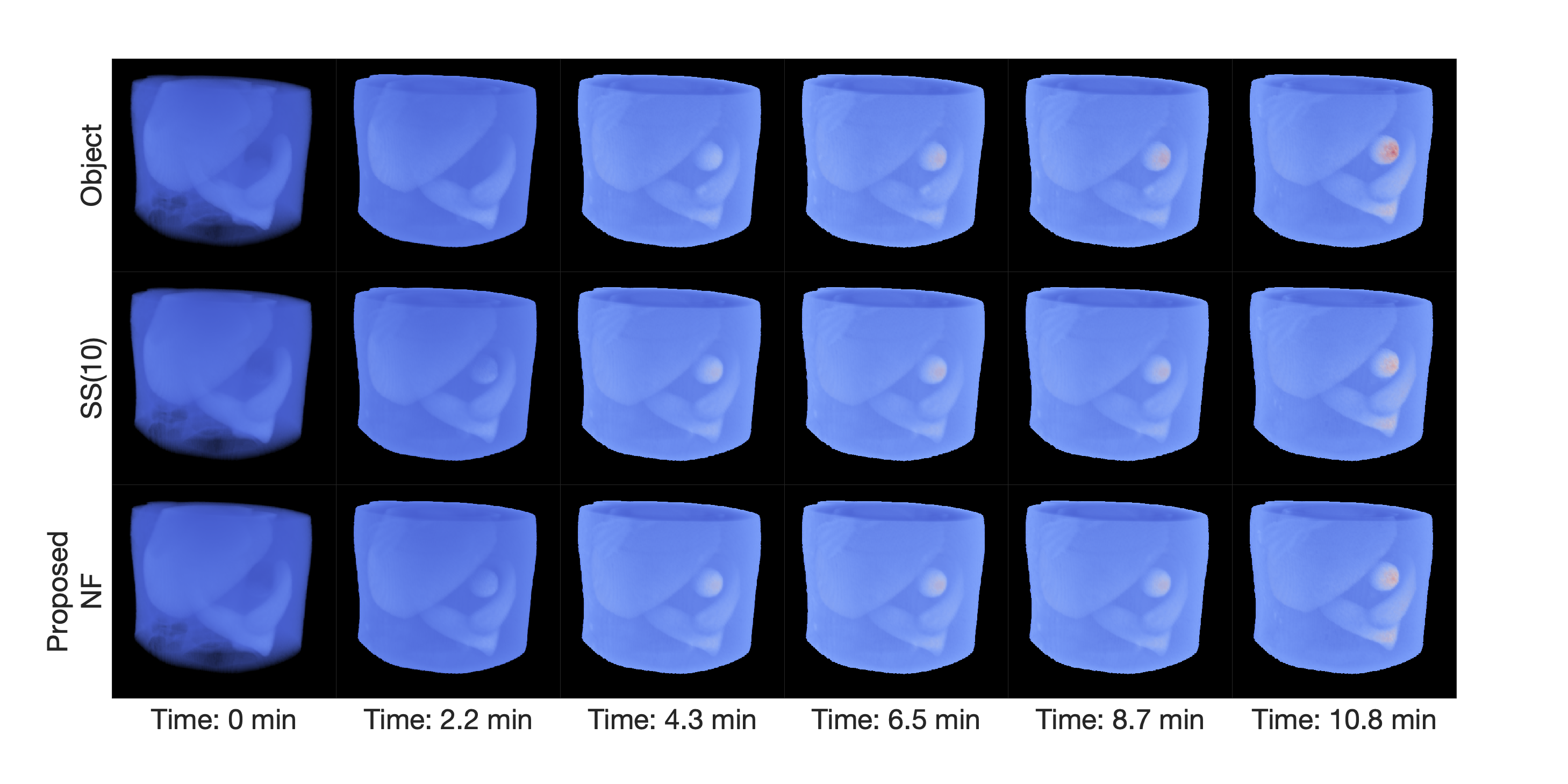}
    \vspace{-5mm}
    \caption{Study 1b: Dynamic embedding in three spatial dimensions. Object (top), Rank 10 semiseparable approximation SS($10$) (second row),  NF representation (bottom row) with proposed POUnet architecture are shown at selected frames throughout the time period. ParaView \cite{ahrens200536} was used for volume rendering. Dynamic range of the rendering is [0, 0.06] using a linear scale. Both methods provide high fidelity representations but slightly underestimate the tumor activity in later time periods. An animation of each representation constructed in this study is available in the supplemental multimedia materials (Video 2).}
    \vspace{-5mm}
    \label{fig:3Drep}
\end{figure*}

\begin{table}[thb]
    \caption{Study 1: Dynamic embedding. RRMSE, SSIM, ROI RRMSE, ROI SSIM, and LAC RRMSE achieved by the NF representations with original and proposed POUnet architectures and the rank $r$ semiseparable approximations SS($r$) in two (2D) and three (3D) spatial dimensions.}
    \label{tab:embeding_acc}
    \centering
    \begin{tabular}{@{}l@{ }|@{}c@{ \ \ }c@{ \ \ }c@{ \ \ }c@{\ \ }c@{ \ \ }c@{}}
    & \# of & RRMSE & SSIM & ROI & ROI & LAC \\ 
    & Params & & & RRMSE & SSIM & RRMSE \\ 
    
    \hline
       SS(12) 2D  & 430 K&$0.0426$ & $0.9964$ & $0.0842$ & $0.9114$ & $0.0282$ \\ 
       Original NF 2D & 425 K  &    $0.2600$ & $0.9399$ & $0.2571$ & $0.8432$ & $0.0591$ \\
       Proposed NF 2D & 427 K & $0.0670$  & $0.9943$ & $0.1005$ & $0.9026$& $0.0291$\\
    \hline
       SS(10) 3D  & 35 M  & $0.0558$ & $0.9983$ & $0.1661$ & $0.9864$ & $0.0988$ \\ 
       Proposed NF 3D &  35 M &   $0.0571$  & $0.9982$ & $0.1661$ & $0.9862$& $0.1174$\\
    \hline
    \end{tabular}
\end{table}

\paragraph{Two spatial dimensions} The accuracy metrics---RRMSE and SSIM over the entire image, RRMSE and SSIM over the region of interest containing the inserted lesion, and the lesion activity curve (LAC) RRMSE---achieved by each representation are reported in Table \ref{tab:embeding_acc}.  The LACs associated with these representations and with the object are shown in Fig. \ref{fig:rep_activity}.  Selected frames of these representations and the object are shown in Fig. \ref{fig:2DRepresentation}. The proposed NF architecture showed superior approximation properties (as quantified by the above accuracy metrics) compared to the original POUnet architecture in \cite{LozenskiNeuralFields}, and comparable accuracy to the semiseparable approximation with a similar number of parameters.

\paragraph{Three spatial dimensions}  The accuracy metrics for the proposed POUnet representation and the semiseparable approximation of rank 10 (SS(10)) are reported in Table \ref{tab:embeding_acc}.  The LACs associated with these representations and with the object are shown in Fig. \ref{fig:3dRepAct}. Selected frames of these representations and the object are shown in Fig. \ref{fig:3Drep}.  The proposed NF architecture achieved comparable accuracy metrics to those achieved by the semiseparable approximation with a similar number of parameters.

\subsection{Study 2: Dynamic reconstruction}
In this study, the proposed ProxNF method was assessed and compared against a well-established spatiotemporal image reconstruction method with nuclear norm regularization (STIR-NN) and, for the problem in two spatial dimensions, with the gradient-based NF method (GradNF) in \cite{LozenskiNeuralFields}. Specifically, a 2D cross-sectional slice and 3D volume extracted from the dynamic tumor perfusion phantom employed in Study 1 were virtually imaged using the stylized PACT system described in Section \ref{subsec:imaging_system}.

\begin{table}[thb]
    \caption{Study 2: Dynamic reconstruction. Number of parameters, RRMSE, SSIM, ROI RRMSE, ROI SSIM, and LAC RRMSE achieved by the proposed ProxNF method, gradient-based NF method (GradNF 2D), and the reference STIR-NN method in two (2D) and three (3D) spatial dimensions.}
    \label{tab:recon_acc}
    \centering
    \begin{tabular}{@{}l@{ }|@{}c@{ \ \ }c@{ \ \ }c@{ \ \ }c@{ \ \ }c@{ \ \ }c@{}}
    & \# of & RRMSE & SSIM & ROI & ROI & LAC \\ 
    & Params & & & RRMSE & SSIM & RRMSE \\ 
    
    \hline
       STIR-NN 2D & 5 M  & $0.2467$ & $0.8822$ & $0.2530$ & $0.7826$ & $0.1406$ \\ 
        GradNF 2D &    427 K &$0.2411$ & $0.8945$ & $0.2469$ & $0.8274$ & $0.0757$ \\
       ProxNF 2D &    427 K &$0.2392$ & $0.9039$ & $0.2404$ & $0.8279$ & $0.0791$ \\
    \hline
       STIR-NN 3D& 52 M  & $0.2915$ & $0.9493$ & $0.4021$ & $0.9085$ & $0.3193$ \\ 
       ProxNF 3D &    35 M & $0.2452$ & $0.9665$ & $0.2288$ & $0.9564$ & $0.1916$ \\
    \hline
    \end{tabular}
\end{table}

\begin{table}[thb]
    \caption{Study 2: Dynamic reconstruction. Number of evaluations of the (randomly sub-sampled) imaging operator (\# evals), batch size (J), and wall-time (W-time).}
\label{tab:comp_cost}
    \centering
\begin{tabular}{@{}l@{ }|@{}r@{ \ \ }r@{ \ \ }l@{ \ \ } }
{} & \# evals & J & W-time\\
\hline
STIR-NN 2D & 140 & 1,296 & 17 CPU minutes\\
GradNF 2D & 128,000 & 324 & 4.0 GPU hours\\
ProxNF 2D & 1,000 & 324 & 1.3 GPU hours\\
\hline
STIR-NN 3D & 1272 & 324 & 110 CPU hours \\
ProxNF 3D & 800 & 324 & 200 GPU hours\\
\hline
\end{tabular}
\end{table}

\begin{figure}
    \centering
    \includegraphics[width = 0.8\columnwidth]{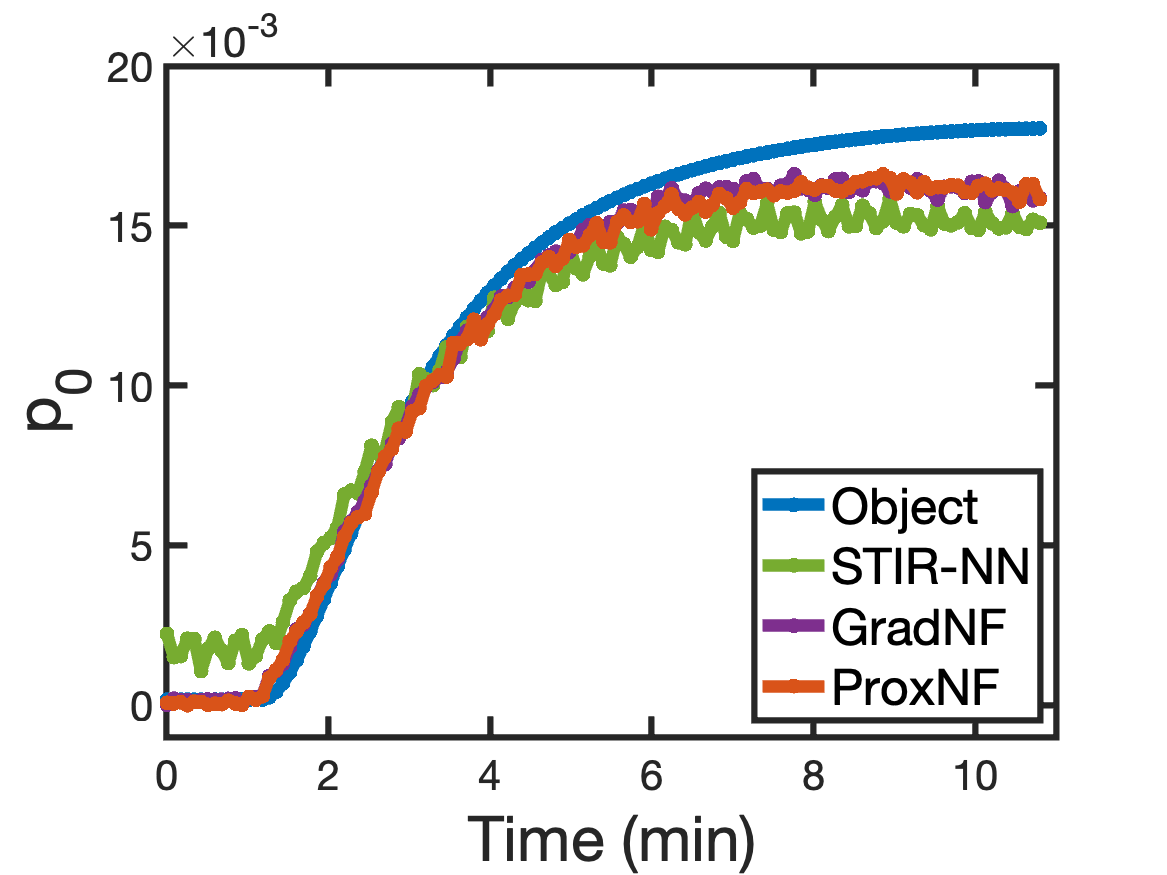}
    \caption{Study 2a: Dynamic reconstruction in two spatial dimensions. LAC of the object (blue), and its estimates obtained by STIR-NN (green), GradNF (purple), and ProxNF (red). Note how the NF estimates are more accurate than the STIR-NN one in the pre- and post-injection phases.}
    \label{fig:crt_activity}
    \vspace{-5mm}
\end{figure}

\begin{figure*}
    \centering

    \includegraphics[width = 0.9\textwidth, trim = {2.5cm 0cm 4cm 1cm},clip]{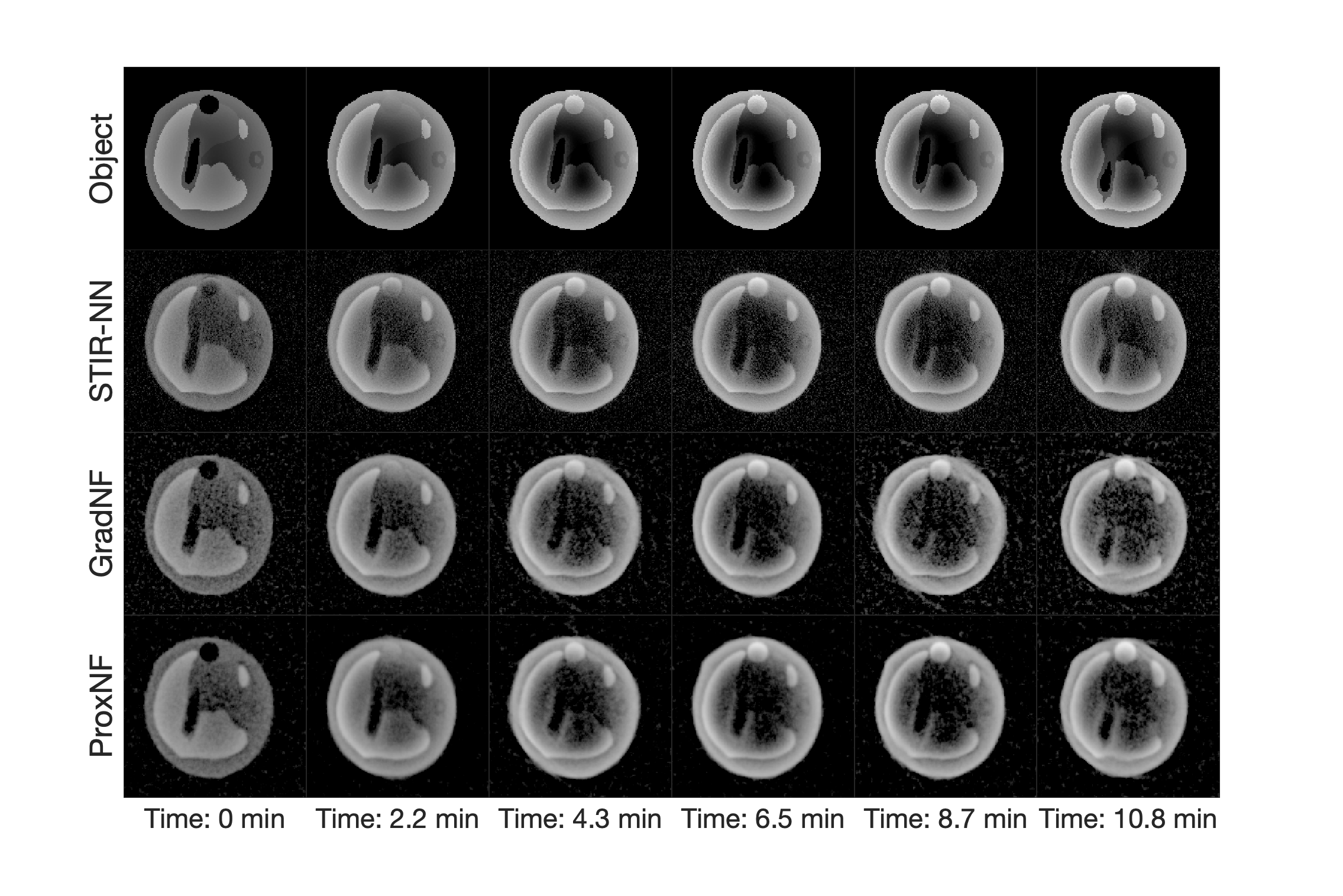}
    \vspace{-10mm}
    \caption{Study 2a: Dynamic reconstruction in two spatial dimensions.  Object (top row) and its estimates produced by STIR-NN (second row), GradNF (third row), and ProxNF (last row) are shown at selected imaging frames. The dynamic range of the grayscale image is $[10^{-4}, 0.06]$. A logarithmic scale was used to increase visualization contrast at depth.  An animation of each reconstructed estimate is available in the supplemental multimedia materials (Video 3). ProxNF can produce an object estimate with comparable accuracy to that produced by STIR-NN  while reducing the memory requirements of one order of magnitude.}
    \label{fig:CRT_recon}
    \vspace{-5mm}
\end{figure*}

\paragraph{Two spatial dimensions} The number of parameters required to represent the reconstructed estimate of the object and the corresponding accuracy metrics---RRMSE and SSIM for the entire image, the RRMSE and SSIM for the region of interest with the inserted lesion, and LAC RRMSE---are reported in Table \ref{tab:recon_acc}.  The ProxNF and GradNF reconstructions achieved similar accuracy and outperformed the STIR-NN reconstruction with respect to all evaluation metrics.  Furthermore, the ProxNF and GradNF reconstructions required less than a tenth of the parameters used by the STIR-NN. The LAC and selected frames of the object and its estimates  are shown in Fig. \ref{fig:crt_activity} and \ref{fig:CRT_recon}, respectively. Table \ref{tab:comp_cost} summarizes the computational cost of the three methods in terms of number of evaluations of the imaging operator and wall-time. Note that, while STIR-NN accessed all $K=1,296$ imaging frames at each iteration to evaluate the imaging operator, ProxNF and GradNF stochastically subsampled the imaging operator using  $J=324$ randomly selected frames at each iteration. While slower than STIR-NN, ProxNF provides significant acceleration ($\sim$128X less evaluations of the imaging operator, $\sim$3X reduction in wall-time) compared to GradNF. 

Additionally, manifold discovery and analysis (MDA) \cite{islam2023mda} was performed on the learned partition network $\boldsymbol{\Psi}_{\boldsymbol{\eta}}$ in Eq. \eqref{eq:mpounet_def} at each ProxNF iteration. MDA provides a two-dimensional, easy-to-visualize representation of a neural network outputs and inner layers. This analysis shows that $\boldsymbol{\Psi}_{\boldsymbol{\eta}}$ quickly stabilizes to a fixed point, thus providing empirical evidence of the ProxNF convergence properties (see Remark \ref{rem:convergence}). A video showing the evolution of the MDA representation of $\boldsymbol{\Psi}_{\boldsymbol{\eta}}$ at each ProxNF iteration is available in the supplementary materials (Video 4).

\begin{figure}
    \centering
    \includegraphics[width = 0.8\columnwidth]{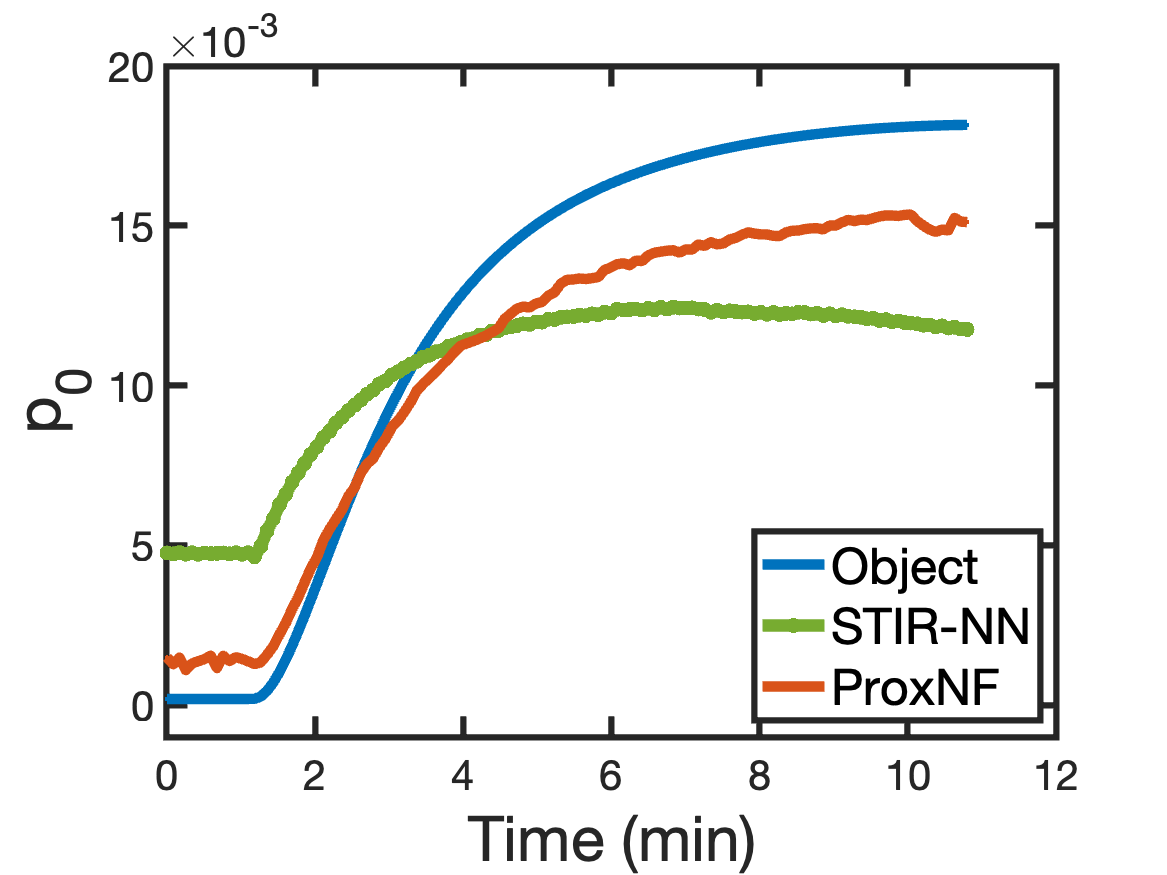}
    \caption{Study 2b: Dynamic reconstruction in three spatial dimensions. LAC of the object (blue) and its estimates produced by STIR-NN (green) and ProxNF (red). The NF representation underestimates the true LAC less severely than STIR-NN.}
    \label{fig:3dRecAct}

\end{figure}

\begin{figure*}
    \centering
    \includegraphics[width = 0.9\textwidth, trim = {2.5cm 0cm 2.5cm 1cm},clip]{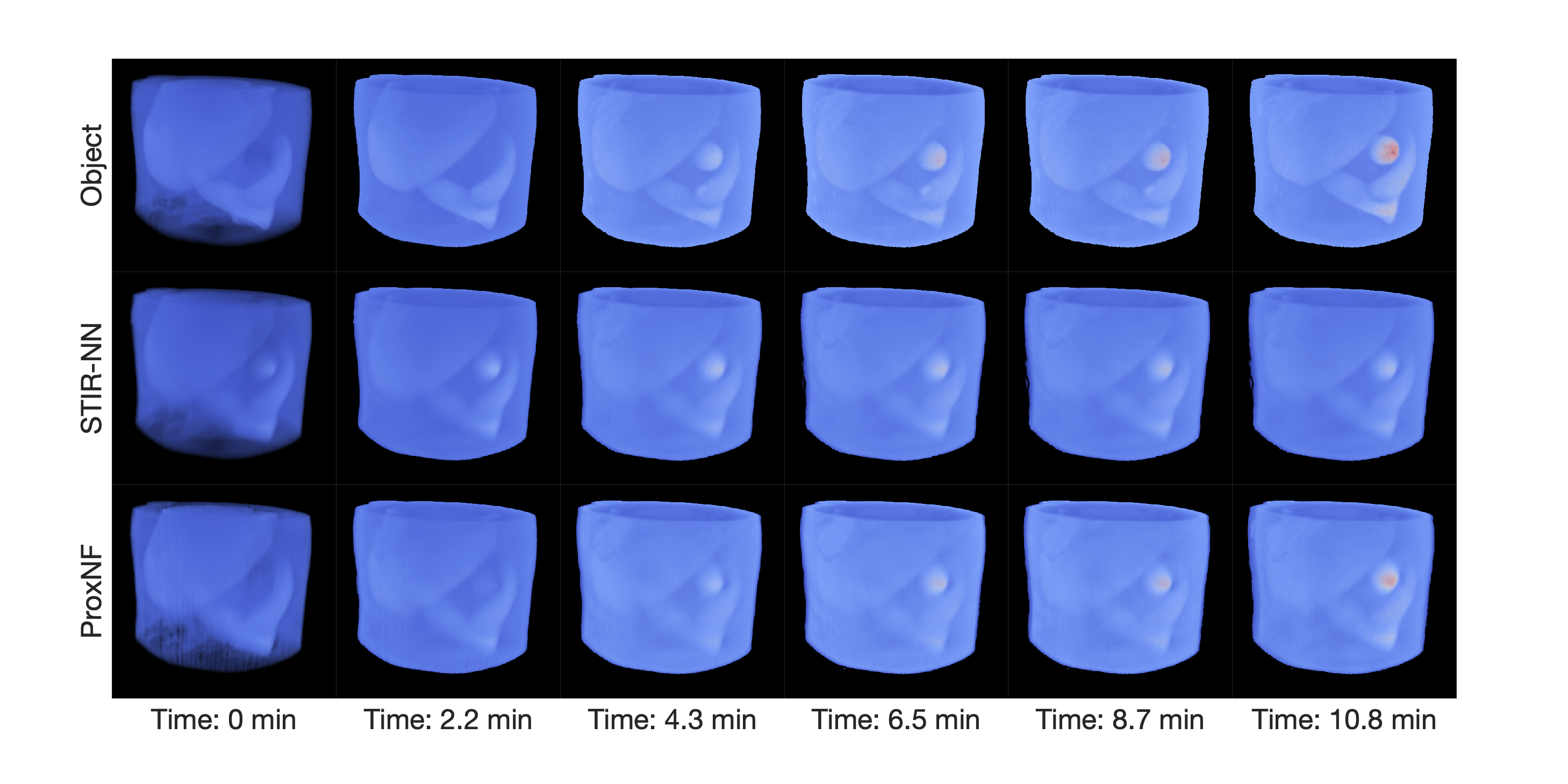}
    \vspace{-5mm}
    \caption{Study 2b: Dynamic reconstruction in three spatial dimensions.
 ParaView  \cite{ahrens200536} was used for volume rendering. 
 Dynamic object (top) and its estimates produced by STIR-NN (middle), and ProxNF (bottom) are shown at selected imaging frames.
 Dynamic range of the rendering is $[0,0.05]$ using a linear scale.  An animation of each reconstruction from this study is available in the supplemental multimedia materials (Video 5). The proposed method can better capture the dynamic changes in the object compared to the reference method (STIR-NN). Furthermore, as shown in Video 4, ProxNF can resolve the respiratory motion of the numerical phantom while the reconstructed estimate produced reference method only captures a time-averaged dynamic over the respiratory cycle.}
    \label{fig:3Drec}
    \vspace{-5mm}
\end{figure*}

\paragraph{Three spatial dimensions} The number of parameters and accuracy metrics for the ProxNF and STIR-NN  methods are reported in Table \ref{tab:recon_acc}. The reconstructed LACs are shown in Fig. \ref{fig:3dRecAct}. Selected frames of these reconstructions and the original object are shown in Fig. \ref{fig:3Drec}. Note that the ProxNF reconstruction is more accurate than the STIR-NN reconstruction with respect to all metrics and requires less evaluations of the imaging operator (c.f. Table \ref{tab:comp_cost}).

\section{Application to in-vivo DCE PACT imaging of tumor perfusion}\label{sec:in-vivo}

The proposed ProxNF method was also demonstrated using retrospective in-vivo DCE PACT imaging data of tumor perfusion in small animal models. Animal studies were conducted in compliance of the current regulations and standards of the National Institutes of Health and approved by the Institutional Animal Care and Use Committee of MD Anderson Cancer Center. In this study, a single cross-sectional slice of a mouse torso bearing a pancreatic cancer xenograft flank tumor
is dynamically imaged at a single light wavelength.

\subsection{Experimental setup and data acquisition}

Measurement data were acquired utilizing the commercially available inVision PACT animal scanner (iThera Medical; Munich Germany). This system employs a toroidally focused, ring-shaped transducer array with 256 detection elements covering 270$^\circ$ around the sample. Transducers' center frequency is 5 MHz and sampling rate 50 MHz. Together with a near-infrared laser (680–980 nm) with a 10 Hz repetition rate, this setup enables cross-sectional tomographic imaging of a small animal torso with an in-plane resolution of 150$\sim$180 $\mu$m. In the experiment, data were continuously acquired at 8 different wavelengths, 6 averages per wavelength before, during and after the injection of the contrast agent. The total duration of the scan was $T=16$ minutes, resulting in 200 imaging frames with a temporal resolution of 4.8 seconds.

\subsection{Image reconstruction studies}

An image reconstruction study was performed utilizing undersampled measurements at a single wavelength (800 nm). Specifically, at each imaging frame, only data collected by 64 transducer elements (8 equispaced groups of 8 contiguous elements each) was employed. The subset of active transducer elements was cyclically rotated for each frame. 
Image reconstruction using the proposed ProxNF method and the STIR-NN method was performed on a $256 \times 256$ spatial grid (square field of view with size 36 mm) and 200 imaging frames.

The NF implemented the same network architecture described in the numerical studies. The nuclear norm reconstruction utilized a maximum rank of 20. 

A reference image was estimated utilizing measurements from all the 256 transducer elements at all time frames by use of the fast spatiotemporal image reconstruction method in \cite{WangXiaetal14}. Based on Morozov discrepancy principle, the rank of the truncated singular value decomposition of the data matrix $\boldsymbol{d}$ was set to 20 in the estimation of the reference image. Accuracy of the ProxNF and STIR-NN reconstructions were then assessed using RRMSE and SSIM, evaluated both on the entire image and on a ROI containing the  lesion, with respect to the reference image. Image quality was also assessed by analyzing the estimated LAC, which is the time-varying spatially-averaged signal over a volume of interest (the segmented lesion).

\subsection{Results and analysis}

\begin{figure}
    \centering
    \includegraphics[width = 0.8\columnwidth]{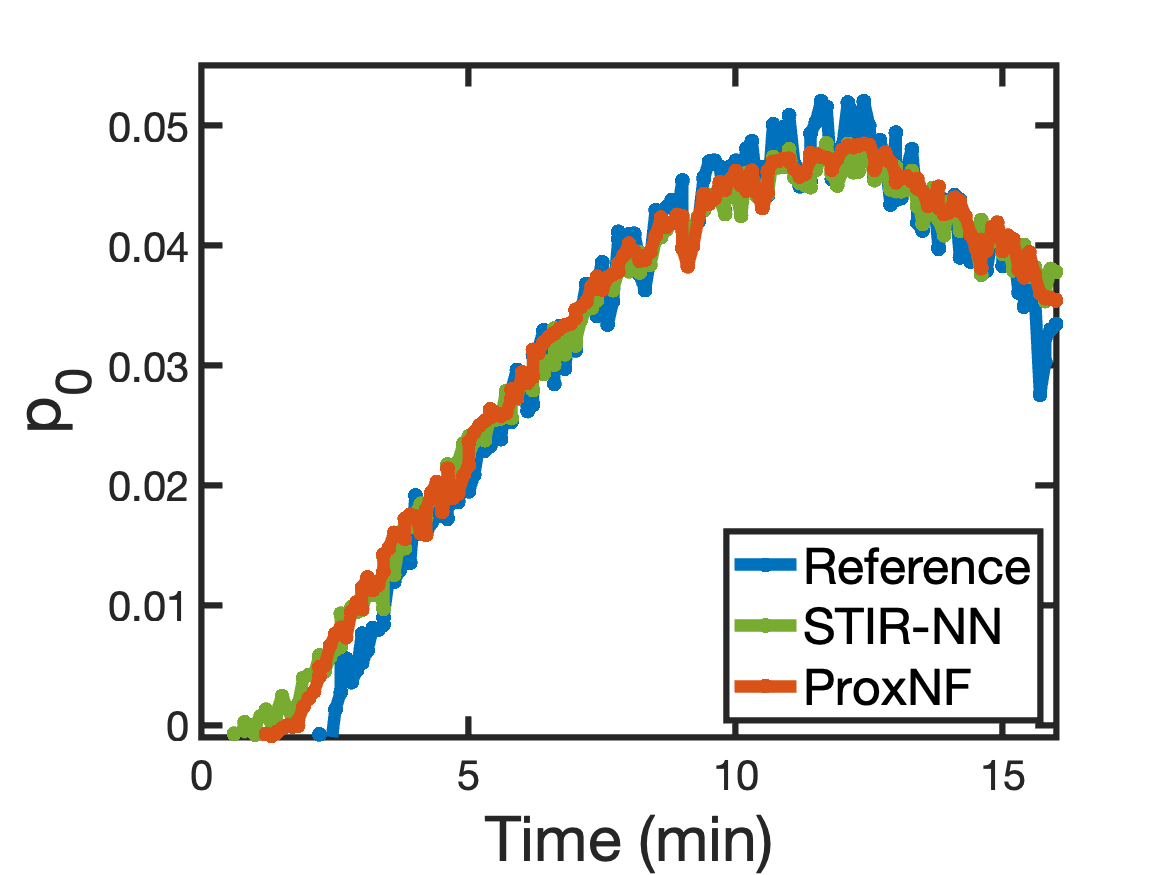}
    \caption{In-Vivo Experiment: LAC of the reference image  (blue), nuclear norm reconstruction (green), and NF reconstruction trained using ProxNF (red). }
    \label{fig:invivo_activity}
        \vspace{-5mm}
\end{figure}

\begin{figure*}
    \centering

    \includegraphics[width = 0.8\textwidth, trim = {2.5cm 0cm 4cm 1cm},clip]{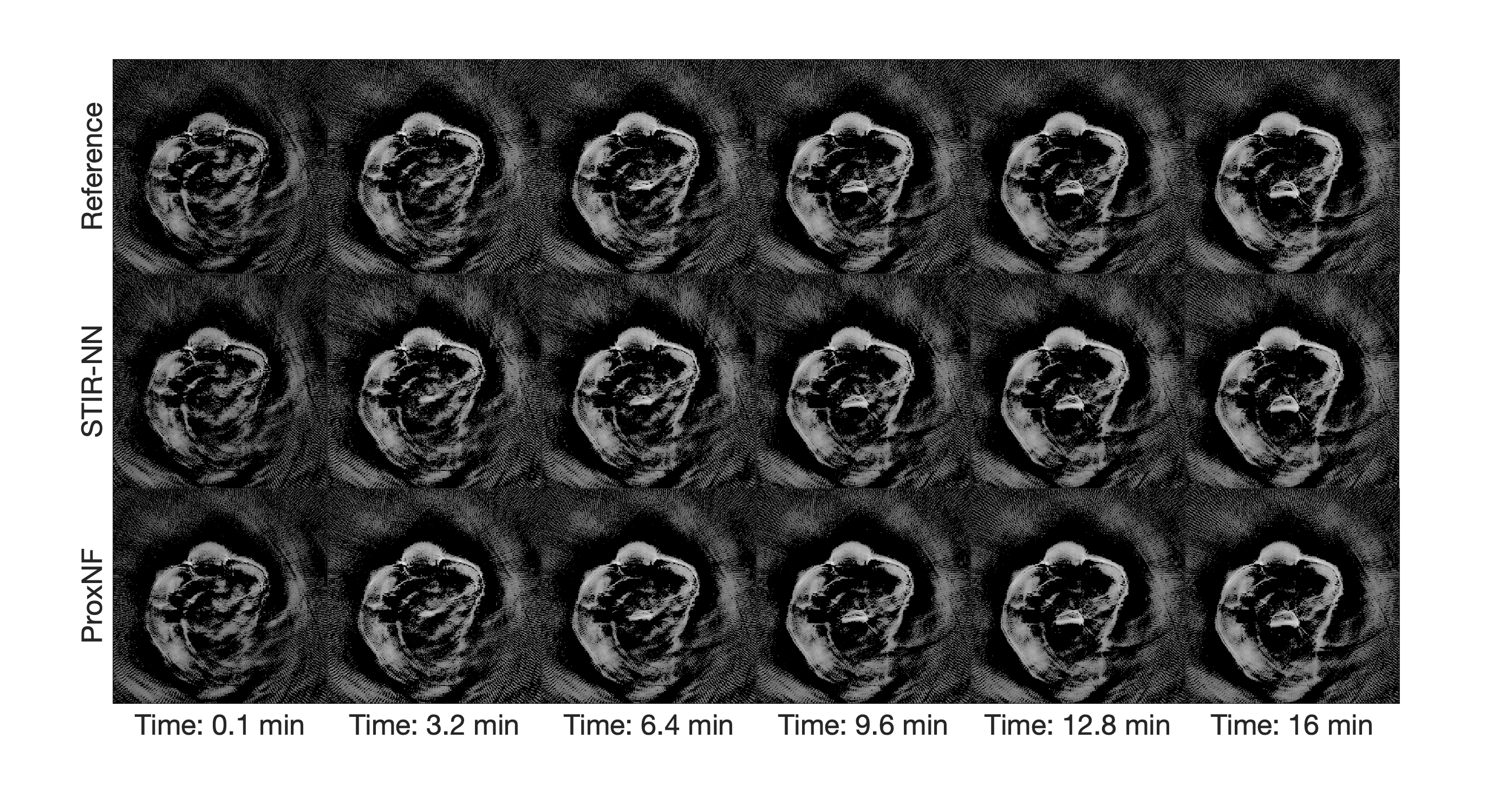}
    \vspace{-6mm}
    \caption{In-Vivo Experiment: Reference image (top) reconstructed utilizing full measurements, STIR-NN (middle), and ProxNF (bottom) reconstructed images from undersampled measurements are shown at selected imaging frames. The dynamic range of the grayscale image is $[10^{-4}, 0.6]$. A logarithmic scale is employed to improve image contrast at depth. An animation of each reconstructed estimate is available in the supplemental multimedia materials (Video 6). ProxNF produces an object estimate with improved accuracy compared to the STIR-NN reconstruction, while halving the memory requirements.}
    \label{fig:InVivo_recon}
    \vspace{-5mm}
\end{figure*}

The number of parameters required to represent the reconstructed image and the corresponding accuracy metrics---RRMSE and SSIM for the entire image, the RRMSE and SSIM for the ROI containing the lesion, and the LAC RRMSE---are reported in Table \ref{tab:invivo_acc}. The ProxNF reconstruction was more accurate compared to the STIR-NN method with respect to all metrics.  Furthermore, the ProxNF reconstruction required half the  number of the parameters used by the STIR-NN. The LAC and selected frames of the reference image and the estimates produced by the ProxNF and STIR-NN methods are shown in Figs. \ref{fig:invivo_activity} and \ref{fig:InVivo_recon}, respectively.

\begin{table}[thb]
    \caption{In-Vivo Experiment: Number of parameters, RRMSE, SSIM, ROI RRMSE, ROI SSIM, and LAC RRMSE achieved by the ProxNF and STIR-NN methods with respect to the reference image}
    \label{tab:invivo_acc}
    \centering
    \begin{tabular}{@{}l@{ }|@{}c@{ \ \ }c@{ \ \ }c@{ \ \ }c@{ \ \ }c@{ \ \ }c@{}}
    & \# of & RRMSE & SSIM & ROI & ROI & LAC \\ 
    & Params & & & RRMSE & SSIM & RRMSE \\ 
    
    \hline
       STIR-NN & 1.31 M  & $0.2441$ & $0.9617$ & $0.2706$ & $0.8379$ & $0.0908$ \\ 
       ProxNF &    730 K &$0.1876$ & $0.9767$ & $0.2449$ & $0.8745$ & $0.0846$ \\
    \hline
    \end{tabular}
\end{table}

\section{Conclusion}\label{sec:conclusion} This work proposes ProxNF, a proximal splitting method for high-resolution, high-dimensional image reconstruction using a neural field (NF) representation. NFs are a special class of neural networks that provide accurate representations of continuous objects and can be utilized for memory-efficient image reconstruction. NFs have demonstrated benefits for several medical imaging applications \cite{molaei2023implicit}, including dynamic imaging\cite{XieTakikawaSaitoEtAl21, dupont2021coin, LozenskiNeuralFields, ReedKimAnirudhetal2021}. The proposed ProxNF method increases the scalability of NF-based image reconstruction approaches enabling high-resolution, long-time horizon volumetric dynamic imaging. ProxNF achieves this by separating computations involving updates of the NF parameters from those involving the imaging operator, thereby reducing the number of evaluations of the imaging operator and allowing for efficient training of NF parameters.
The efficacy of the proposed partition-of-unity-based NF architecture and ProxNF method were illustrated in two numerical studies and in-vivo  dynamic contrast-enhanced photoacoustic computed tomography (DCE PACT) imaging of tumor perfusion in small animal models. 

The first numerical study showed that the proposed partition of unity network (POUnet) architecture is able to accurately represent a dynamic image in two and three spatial dimensions, achieving an accuracy comparable to that yielded by a space-time semiseparable approximation using a similar number of parameters. 

The second numerical study considered dynamic image reconstruction problems in two and three spatial dimensions. A numerical tumor perfusion phantom was virtually imaged with a limited number of views per imaging frame. This study demonstrated that the NF architecture trained using ProxNF outperforms a well-established image reconstruction method utilizing nuclear norm regularization (STIR-NN) in terms of RRMSE and SSIM while using a smaller number of parameters to represent the sought-after image. Furthermore, the NF method yielded a more accurate estimate of the lesion activity curve compared to that achieved by the method using nuclear norm regularization. 

In the in-vivo DCE PACT study, an axial cross-section of a mouse was estimated utilizing undersampled measurements at each imaging frame. The accuracy of the proposed method was assessed utilizing a reference image reconstructed from the fully sampled measurements, and compared to that achieved by STIR-NN. In this application, ProxNF achieved a higher accuracy, as well as a 50\% reduction in the number of parameters, compared to STIR-NN. 

In summary, this work established the feasibility of NF representations for dynamic image reconstruction of high-resolution, high-dimensional images. ProxNF, the proposed proximal splitting method, helps overcome the computational shortcomings of previous approaches to spatiotemporal image reconstruction using NFs \cite{LozenskiNeuralFields}. A key advantage of the proposed method is that it is agnostic of the chosen neural representation. The plug-and-play formulation of ProxNF allows to seamlessly replace the POUnet architecture used in this work with any other state-of-the-art NF architecture\cite{SitzmannMartelBergmanetal2020, dupont2021coin, MartelLindellLinetal2021, XieTakikawaSaitoEtAl21, LiuSunetal2021, LozenskiNeuralFields}. Additionally, ProxNF could easily be generalized to other approaches for representing images with neural networks, such as deep image priors \cite{ulyanov2018deep}, and still provide a separation of concerns between optimizing network parameters and evaluations of the imaging operator.  

The primary limitation of this work is the relatively long computational time compared to well-established image reconstruction method utilizing nuclear norm regularization. However, the use of high performance deep-learning backends, such as \texttt{JAX}\cite{frostig2018compiling} and improved NF embedding algorithms can potentially mitigate this issue. Future work may extend the proposed method to 5D (three spatial dimensions, one time dimension, and one wavelength dimensions) quantitative PACT, in which dynamic changes in an object's chromophore concentrations are estimated from multispectral PACT data by use of a non-linear multiphysics imaging operator coupling photon transport and wave propagation.

\appendices
\section{Code and Data Availability}

The Python code implementing the proposed dynamic image reconstruction method using neural field is available from \cite{Lozenski2022code} under GPLv3. The absorption coefficient and induced pressure maps of the dynamic object used in the numerical studies are available from \cite{Lozenski2023phantom} under CC-0 public dedication.

\bibliographystyle{IEEEtran}
\bibliography{local, references}

\end{document}